\documentclass{article}

\newcommand\Real{\mbox{Re}}
\newcommand\Imag{\mbox{Im}}
\newcommand\diag{\mbox{diag}}

\newcommand{\expo}{\,\mathrm{e}}
\newcommand{\cvar}{\,\mathrm{i}}
\newcommand{\ud}[1]{\,\mathrm{d}#1}
\newcommand{\besselj}{\mathrm{J}}
\newcommand{\besselh}{\mathrm{H}}
   
\usepackage{amsmath,amsfonts,graphicx}

\begin{document}

\title{Modelling wave-induced sea ice breakup in the marginal ice zone}

\author{
F. Montiel$^{1}$ and V. A. Squire$^{1}$}

\date{$^{1}$Department of Mathematics and Statistics, University of Otago, PO Box 56, Dunedin, New Zealand\\}

%

\maketitle 

\begin{abstract}

\noindent A model of ice floe breakup under ocean wave forcing in the marginal ice zone (MIZ) is proposed to investigate how floe size distribution (FSD) evolves under repeated wave breakup events. 
A three-dimensional linear model of ocean wave scattering by a finite array of compliant circular ice floes is coupled to a flexural failure model, which breaks a floe into two floes provided the two-dimensional stress field satisfies a breakup criterion. 
A closed-feedback loop algorithm is devised, which (i)~solves wave scattering problem for a given FSD under time-harmonic plane wave forcing, (ii)~computes the stress field in all the floes, (iii)~fractures the floes satisfying the breakup criterion and (iv)~generates an updated FSD, initialising the geometry for the next iteration of the loop.
The FSD after 50 breakup events is uni-modal and near normal, or bi-modal. 
Multiple scattering is found to enhance breakup for long waves and thin ice, but to reduce breakup for short waves and thick ice.
A breakup front marches forward in the latter regime, as wave-induced fracture weakens the ice cover allowing waves to travel deeper into the MIZ.

\end{abstract}


\section{Introduction}\label{sec:Intro}

The Arctic marginal ice zone (MIZ) that separates open ocean from the interior pack ice is experiencing rapid changes as a result of high latitude climate change.
During summer, for example, its extent relative to total sea ice area is expanding \cite{strong_rigor13}, suggesting an increasing presence of thinner, loosely packed ice floes.
Changes in environmental forcings, e.g.\ heat, winds and ocean waves, acting in partnership with positive feedback processes, are responsible for this transformation. 
Ocean waves, in particular, have been observed to break up the sea ice under flexural failure and therefore to contribute to the increasing extent of the MIZ, which is, in turn, more sensitive to summer melting because of the increased total perimeter of the ice floes created \cite{perovich_jones14}. 

\noindent Correspondingly, sea ice loss increases open water extent and allows for more energetic swell to develop in the Arctic Basin \cite{Thomson_rogers14,thomson_etal16}, with the potential to fracture the sea ice further and cause additional melting.
Although indirect observational evidence of this positive feedback mechanism was recently proposed by \cite{kohout_etal14} in the Antarctic MIZ, its impact on sea ice extent has not been quantified.
Modelling the two-way coupling between the wave and sea ice systems on oceanic scales is needed to remedy this shortcoming.

The vast majority of modelling studies on ocean wave interactions with sea ice have attempted to quantify wave attenuation and directional spreading as a result of scattering \cite{squire11,bennetts_squire12a,zhao_shen16,montiel_etal16} and dissipation \cite{mosig_etal15,squire_montiel16} by the constituent ice floes, within the scope of linear water wave theory.
These effects have recently been parametrised in spectral wave models, e.g.\ WAVEWATCH III$^{\textrm{\textregistered}}$, to complement the description of physical processes influencing ocean wave propagation on a global scale and assess the role sea ice has on wave climate in the polar seas \cite{doble_bidlot13,collins_etal15,li_etal15,ardhuin_etal16,rogers_etal16}, acknowledging that the validity of such parametrisations is still the subject of much current research. 

In contrast, very little is known about the impact of ocean waves on the breakup of sea ice floes in the MIZ.
Observations have shown that floe size distribution (FSD), defined as the statistical distribution of floe sizes (e.g.\ mean caliper diameter or diameter of a circular floe with the same surface area) in the MIZ, satisfies a split power law with different regimes associated with different values of the power exponent \cite{rothrock_thorndike84,toyota_etal06,toyota_etal11,perovich_jones14}.
It is unclear, however, how waves contribute to the emergence of this power law, as flexural failure is not expected to occur below a critical floe size of order O(10\,m) \cite{mellor86}, which would suggest that the probability density function of the FSD should decrease to zero as floes become small unless other breakup mechanisms are imposed.
In this study, we address this question by modelling the breakup of an ice cover under a sustained wave event, with the goal of establishing the FSD emerging from wave forcing alone, i.e.\ isolated from wind, collisions and any other sources of sea ice breakup.

Very few models have attempted to describe the breakup of sea ice in the MIZ due to waves alone. 
Dumont et al.\ \cite{dumont_etal11} were the first to propose a numerical model for the transport of ocean waves in the MIZ due to scattering by the constituent ice floes, in which a parametrisation of ice floe breakup was included. 
In each cell of the discretised spatial domain, the FSD was described by a power law and parametrised by its minimum, mean and maximum floe size. 
At each time step, the FSD was updated according to a breakup criterion (discussed later) depending on wave amplitude in the cell and floes repeatedly fracturing in half with a prescribed probability, preserving the power law distribution.
Williams et al. \cite{williams_etal13a,williams_etal13b} extended the work of \cite{dumont_etal11} by considering a more realistic breakup criterion, but used the same parametrisation of the FSD.
These authors focused their analysis on estimating the maximum distance from the ice edge where breakup can take place, which they define as the MIZ width, and did not examine the evolution of the power law FSD during the breakup process.
Implementation of two-way coupling between large scale sea ice models, e.g.\ CICE or neXtSIM, and spectral wave models based on these modelling approaches is currently being investigated \cite{bennetts_etal17,williams_etal17}.

Other modelling studies have considered the evolution of the FSD in the framework of large scale sea ice models. 
Zhang et al. \cite{zhang_etal15} proposed a continuum transport equation for the probability density function (PDF) of the FSD, extending a similar approach for ice thickness distribution used for describing ice thickness in large scale sea ice models.
The FSD is advected in time and space subject to a prescribed horizontal ice velocity field and a number of sources and sinks that describe the effects of thermodynamics (melting and freezing), lead opening, ridging and fragmentation, noting the latter is parametrised in a highly simplified manner (uniformly redistributed floe sizes).
Horvat and Tziperman \cite{horvat_tziperman15} extended the work of \cite{zhang_etal15} to the joint floe size and thickness distribution and enhanced the parametrisation of source and sink terms, particularly wave-induced floe fracture which accounts for the strain field generated by a wave spectrum.
Also note the stochastic model of FSD evolution based on the generalised Lotka-Volterra equation \cite{herman10}, which exhibits stable solutions that approximate observed FSDs well.

Here, the three-dimensional phase-resolving scattering model of wave energy attenuation in the MIZ reported by Montiel et al.\ \cite{montiel_etal16} is enhanced with a floe breakup model, allowing us to investigate the two-way wave--MIZ coupling in an idealised setting. 
The MIZ is constructed as an array of circular elastic floes with prescribed FSD and the forcing field is approximated by a monochromatic plane wave. 
Solution to the wave interaction problem provides a full description of (i)~the wave field throughout the MIZ and (ii)~the bending experienced by each floe.
The latter information is used to derive a measure of elastic deformation in each floe which, if larger than a critical value, results in floe fracture.
The post-breakup updated FSD is then fed back into the geometrical description of the MIZ, leading to a new solution of the wave interaction, which is in turn used to approximate ice floe breakup.
Running this feedback loop simulation a sufficient number of times, we reach a steady state FSD, which depends on the ice and wave parameters.
The main goals of this investigation are (i) to study the evolution of the FSD towards its steady state under repeated wave breakup events, (ii) to determine the effect of multiple scattering on the steady state FSD and (iii) to examine the dependence of the FSD on the ice and wave parameters of this model.

A key novel feature of the flexural failure model proposed here is that it accounts for the two-dimensional stress field defined over the surface of each deformed floe.
It is based on the two-dimensional Mohr-Coulomb (MC) stress criterion, which assesses mechanical failure from the combined level of tensile and compressive deformations at each point of the floe.
The MC stress criterion has been used to estimate fracture in the elasto-brittle rheological sea ice model neXtSIM under horizontal deformations and wave-induced flexure \cite{williams_etal17}.
The latter fracture model is a simplified one-dimensional version of the MC criterion, as wave-induced ice flexure is approximated using an elastic beam model of ice floe.
This is in line with previous flexural failure models \cite{kohout_meylan08,dumont_etal11,williams_etal13a,williams_etal13b,horvat_tziperman15,kohout_etal16}, in which deformation at each point of the beam is simply quantified by the curvature of its vertical displacement function.
To our knowledge, the wave-induced breakup model of ice floes considered here is the first one to account for the additional spatial dimension.

We do not attempt to compare our model with experimental measurements in this investigation. 
Ice floe breakup by ocean waves in the MIZ has been reported via either in-situ, e.g.\ \cite{liu_mollo-christensen88,prinsenberg_peterson11,collins_etal15,kohout_etal16}, or remote sensing observations, e.g.\ \cite{holt_martin01,doble_bidlot13,wang_etal16}.
These papers describe qualitative or quantitative changes in the FSD after a large wave event breaks up the MIZ.
It is unclear, however, which physical processes have contributed to the observed changes in FSD, as it is not possible to isolate the effect of waves.  
For this reason, we focus our analysis on gaining a theoretical understanding of how ocean waves may influence sea ice breakup in the MIZ and the associated FSD. 

\section{Preliminaries}\label{sec:Prelim}

We consider a three-dimensional seawater domain with constant finite depth, $h$ say, and infinite horizontal extent. 
Cartesian coordinates $\mathbf{x}=(x,y,z)$ are used to locate points in the water domain, such that the planes $z=0$ and $z=-h$ coincide with the unperturbed free surface and the flat impermeable seabed, respectively.
The seawater is approximated as an inviscid and homogeneous incompressible fluid with density $\rho_0\approx1025\,\text{kg\,m}^{-3}$.

A finite array of $N_f$ compliant sea ice floes is assumed to be freely floating at the equilibrium surface of the water domain.
Each floe is circular with uniform thickness $D$ and Archimedean draught $d=(\rho/\rho_0)D$, where $\rho\approx922.5\,\text{kg\,m}^{-3}$ is the density of sea ice.
The radius of floe $i$ is $a_i$ and its centre has coordinates in the horizontal plane $(x_i,y_i)$.
The horizontal region of seawater covered by a floe is denoted by
\begin{equation}\label{eq:IceCoveredRegion}
\Omega_i = \left\{(x,y)\,\in\,\mathbb{R}^2:\,(x-x_i)^2+(y-y_i)^2 \le a_i^2\right\}
\end{equation}
for any $i\,\in\,I$, where $I=\{1,2,\ldots,N_f\}$. 
We further denote their union as $\Omega=\Omega_1\,\cup\,\Omega_2\,\cup\,\ldots\,\cup\,\Omega_{N_f}$ and the horizontal region covered by a free surface as $\Omega_0=\mathbb{R}^2\,\setminus\,\Omega$.

We consider time-harmonic perturbations in the water with prescribed radian frequency $\omega$.
Assuming the flow is irrotational, the velocity field in the water domain is expressed as $(\nabla,\partial_z)\Real\{(g/\cvar\omega)\phi(\mathbf{x})\expo^{-\cvar\omega t}\}$, where $\nabla=(\partial_x,\partial_y)$ and $g\approx9.81\rm\,m\,s^{-1}$ is the acceleration due to gravity. 
The complex-valued potential field $\phi$ then satisfies Laplace's equation in the water domain
\begin{equation}\label{eq:Laplace}
	\nabla^2\phi+\partial_z^2\phi = 0 \quad \text{for} \quad \mathbf{x}\,\in\,\big(\Omega_0\times(-h,0)\big)\,\cup\,\big(\Omega \times(-h,-d)\big).
\end{equation}
The condition of no normal flow on the seabed yields, in addition, the Neumann boundary condition
\begin{equation}\label{eq:Bottom}
\partial_z\phi = 0 \quad \text{on} \quad z=-h.
\end{equation}

We assume that the perturbations in the water induce a flow characterised by a vertical displacement at the free surface that is small compared to the horizontal characteristic length of the flow.
The linearised free surface boundary condition then takes the form
\begin{equation}\label{eq:FreeSurface}
\partial_z\phi = \alpha\phi \quad \text{on} \quad z=0 \quad \text{for} \quad (x,y)\,\in\,\Omega_0,
\end{equation}
where $\alpha=\omega^2/g$.

We introduce a coupling between the vertical deformations experienced by each ice floe and the flow in the water domain.
Horizontal motions of the floes are neglected.
We model the ice floe vertical deformations using Kirchhoff-Love, thin-elastic plate theory.
This model is valid provided (i) ice floe diameters are large compared to the thickness, and (ii) vertical deformations are small compared to the thickness.
The boundary condition on the underside of the ice floes is then given by 
\begin{equation}\label{eq:ElasticPlate}
\left(F\nabla^4 + \rho_0 g - \rho D \omega^2 \right)\partial_z\phi = \rho_0\omega^2\phi \quad \text{on} \quad z=-d \quad \text{for} \quad (x,y)\,\in\,\Omega,
\end{equation}
where $F=ED^3/12(1-\nu^2)$ is the flexural rigidity of the floe, which depends on the effective flexural modulus $E$ \cite{mellor86} and Poisson's ratio $\nu$. 
The values $E\approx6\,\text{GPa}$ and $\nu\approx0.3$ are commonly used for sea ice.

The requirement that each floe $i\,\in\,I$ has no horizontal motion is written
\begin{equation}\label{eq:NoHorizMotion}
\partial_{r_i}\phi = 0 \quad \text{on} \quad r_i=a_i \quad \text{for} \quad -d<z<0
\end{equation}
and the free edge conditions of zero bending moment and zero vertical shear stress at the edge are respectively
\begin{equation}\label{eq:NoBendEdge}
\left[r_i^2\nabla_{r_i,\theta_i}^2 - (1-\nu)\left(r_i\partial_{r_i} + \partial_{\theta_i}^2\right)\right]\partial_z\phi = 0 \quad \text{on} \quad (r_i,z)=(a_i,-d),
\end{equation}
and 
\begin{equation}\label{eq:NoShearEdge}
\left[r_i^3\partial_{r_i}\nabla_{r_i,\theta_i}^2 + (1-\nu)\left(r_i\partial_{r_i} -1\right)\partial_{\theta_i}^2\right]\partial_z\phi = 0 \quad \text{on} \quad (r_i,z)=(a_i,-d),
\end{equation}
where $(r_i,\theta_i)$ are the local polar coordinates with origin at the centre of floe $i$, defined by $(x-x_i,y-y_i)=(r_i\cos\theta_i,r_i\sin\theta_i)$. 
The operator $\nabla_{r_i,\theta_i}=(\partial_{r_i}+1/r_i,(1/r_i)\partial_{\theta_i})$ has also been introduced.

An ambient flow in the water domain is prescribed with potential
\begin{equation}\label{eq:IncWave}
\phi^{\mathrm{am}}(\mathbf{x}) = \psi_0(z)\int_{-\pi/2}^{\pi/2}A^{\mathrm{am}}(\tau)\expo^{\cvar k_0(x\cos\tau+y\sin\tau)}\ud{\tau},
\end{equation}
which satisfies \eqref{eq:Laplace}--\eqref{eq:FreeSurface} and forces a non-trivial solution to the boundary-value problem \eqref{eq:Laplace}--\eqref{eq:NoShearEdge}.
Equation \eqref{eq:IncWave} defines the coherent superposition of plane waves travelling in the positive $x$-direction at the surface of an open ocean and with amplitudes depending continuously on the propagation angle $\tau$ with respect to the $x$-axis.
The flow in the vertical direction is described by the function $\psi_0(z)=\cosh k_0(z+h)/\cosh k_0h$, where $k_0$ denotes the wavenumber of travelling waves in the open ocean (defined later).

We seek a solution of the boundary-value problem \eqref{eq:Laplace}--\eqref{eq:NoShearEdge} of the form $\phi = \phi^{\mathrm{am}} + \phi^{\mathrm{S}}$, where $\phi^{\mathrm{S}}$ is the potential of the scattered wave field due to the presence of the ice floes in response to the ambient wave potential $\phi^{\mathrm{am}}$. 
In the far field, the scattered wave potential satisfies the Sommerfeld radiation condition
\begin{equation}
\sqrt{r}\left(\partial_r - \cvar k\right)\phi^{\mathrm{S}} \to 0 \quad \text{as} \quad r \to \infty,
\end{equation}
where $r=\sqrt{x^2+y^2}$. 

\section{Wave scattering model}\label{sec:ScatModel}

\subsection{Single floe scattering}\label{sec:SingleFloeScat}

We decompose the potential in the exterior open water region adjacent to any floe $i\,\in\,I$ (i.e.\ for $r_i > a_i$) as $\phi\equiv\phi^{(i)}_{\mathrm{ext}} = \phi^{(i)}_{\mathrm{in}} + \phi^{(i)}_{\mathrm{sc}}$, where $\phi^{(i)}_{\mathrm{in}}$ is the local incident wave potential generated by sources away from the floe and $\phi^{(i)}_{\mathrm{sc}}$ is the scattered wave potential generated due to the presence of floe $i$. 
Standard cylindrical eigenfunction expansions are used to express these potentials (see, e.g., \cite{peter_etal03})
\begin{subequations}\label{eq:EigFunExp}
\begin{equation}\label{eq:EigFunExpIn}
\phi^{(i)}_{\mathrm{in}}(\mathbf{x}) = \sum_{m=0}^{\infty} \psi_m(z) \sum_{n=-\infty}^{\infty} a_{m,n}^{(i)} \besselj_n(k_mr_i)\expo^{\cvar n\theta_i} \quad \text{for} \quad \mathbf{x}\,\in\,\Omega_0\times(-h,0)
\end{equation}
and
\begin{equation}\label{eq:EigFunExpSc}
\phi^{(i)}_{\mathrm{sc}}(\mathbf{x}) = \sum_{m=0}^{\infty} \psi_m(z) \sum_{n=-\infty}^{\infty} b_{m,n}^{(i)} \besselh_n(k_mr_i)\expo^{\cvar n\theta_i}  \quad \text{for} \quad \mathbf{x}\,\in\,\Omega_0\times(-h,0).
\end{equation}
The potential $\phi\equiv\phi^{(i)}_{\mathrm{int}}$ in the interior region to floe $i$ is expanded as
\begin{equation}\label{eq:EigFunExpInt}
\phi^{(i)}_{\mathrm{int}}(\mathbf{x}) = \sum_{m=-2}^{\infty} \zeta_m(z) \sum_{n=-\infty}^{\infty} c_{m,n}^{(i)} \besselj_n(k_mr_i)\expo^{\cvar n\theta_i} \quad \text{for} \quad \mathbf{x}\,\in\,\Omega_i\times(-h,-d).
\end{equation}
In these expansions, $\besselj_n$ and $\besselh_n$ denote the Bessel and Hankel functions of the first kind of order $n$, respectively.
\end{subequations}
The eigenfunctions describing the fluid flow in the vertical direction in the open water and ice-covered regions are given by 
\[\psi_m(z)=\frac{\cosh k_m(z+h)}{\cosh k_m h}, \quad m\ge0, \quad \text{and} \quad \zeta_m(z)=\frac{\cosh \kappa_m(z+h)}{\cosh \kappa_m (h-D)}, \quad m\ge-2, \]
respectively.

The quantities $k_m$, $m\ge0$, are solutions of the open water dispersion relation
\begin{equation}\label{eq:DispRelOW}
k\tanh kh = \alpha.
\end{equation}
We denote by $k_0$ the only positive real root of \eqref{eq:DispRelOW}, and by $k_1, k_2, k_3, \ldots$ the infinite number of purely imaginary roots with positive imaginary part ordered such that $\Imag(k_{m+1})>\Imag(k_{m})$ for all $m\ge1$. 
The real root $k_0$ is the wavenumber of horizontally travelling wave modes at the free surface of the open water region, while the imaginary roots $k_m$, $m\ge1$, are associated with horizontally evanescent wave modes decaying exponentially faster from their source for increasing values of $m$.

The quantities $\kappa_m$, $m\ge-2$, introduced in \eqref{eq:EigFunExpInt} are solutions of the ice-covered dispersion relation 
\begin{equation}\label{eq:DispRelIce}
(\beta\kappa^4 + 1 - \alpha d)\kappa\tanh \kappa (h-d) = \alpha.
\end{equation}
The scaled elastic constant $\beta=F/\rho_0 g$ has been introduced in \eqref{eq:DispRelIce}.
We denote by $\kappa_0$ the only positive real root of \eqref{eq:DispRelIce}, which is the wavenumber associated with horizontally travelling wave modes at the water-ice interface. 
In addition, $\kappa_1, \kappa_2, \kappa_3, \ldots$ designate the infinitely many imaginary roots with increasingly large imaginary parts associated with evanescent waves, and $\kappa_{-2}$ and $\kappa_{-1}$ denote the two remaining complex roots with positive imaginary part that are associated with damped travelling wave modes.

A relationship exists between the unknown coefficients $a_{m,n}^{(i)}$, $b_{m,n}^{(i)}$ and $c_{m,n}^{(i)}$ of the eigenfunction expansions given in \eqref{eq:EigFunExp}, as a consequence of the boundary conditions prescribed on the surface $r_i=a_i$. 
In addition to the condition of no horizontal motions \eqref{eq:NoHorizMotion}, continuity of fluid pressure and normal velocity is imposed at the interface between the open water and ice-covered regions, i.e.\
\begin{equation}
\phi^{(i)}_{\mathrm{ext}} = \phi^{(i)}_{\mathrm{int}} \quad \text{and} \quad \partial_{r_i}\phi^{(i)}_{\mathrm{ext}} = \partial_{r_i}\phi^{(i)}_{\mathrm{int}} \quad \text{on} \quad r_i=a_i \quad \text{for} \quad -h<z<-d.
\end{equation}

A numerical solution is obtained by approximating the series expansions in \eqref{eq:EigFunExp} as partial sums, such that $m \le M$ and $|n| \le N$, where $M$ and $N$ are convergence parameters. 
We apply the eigenfunction matching method (EMM) proposed by \cite{montiel_etal13b}.
It provides the following matrix relations between the coefficients
\begin{equation}\label{eq:DTMs}
\mathbf{b}^{(i)} = \mathbf{S}_{\mathrm{ext}}^{(i)} \mathbf{a}^{(i)} \quad \text{and} \quad \mathbf{c}^{(i)} = \mathbf{S}_{\mathrm{int}}^{(i)} \mathbf{a}^{(i)},
\end{equation}
where $\mathbf{a}^{(i)}$, $\mathbf{b}^{(i)}$ and $\mathbf{c}^{(i)}$ are the column vectors containing the coefficients $a_{m,n}^{(i)}$, $b_{m,n}^{(i)}$ and $c_{m,n}^{(i)}$. 
The  vectors $\mathbf{a}^{(i)}$ and $\mathbf{b}^{(i)}$ have dimension $(M+1)(2N+1)$ while $\mathbf{c}^{(i)}$ has dimension $(M+3)(2N+1)$.
Therefore the matrices $\mathbf{S}_{\mathrm{ext}}^{(i)}$ and $\mathbf{S}_{\mathrm{int}}^{(i)}$ have dimensions $((M+1)(2N+1))^2$ and $(M+3)(2N+1)\times(M+1)(2N+1)$, respectively, and are referred to as the exterior and interior diffraction transfer matrices (DTMs).
The DTMs describe the scattering properties of each floe.

\subsection{Multiple scattering}\label{sec:MultiFloeScat}

We use a self-consistent method to resolve wave interactions with the array of $N_f$ ice floes, in which the incident field $\phi^{(i)}_{\mathrm{in}}$ on each floe $i\,\in\,I$ is given by the coherent superposition of the prescribed ambient field and the field scattered by all the other floes.
This is expressed as
\begin{equation}
\phi^{(i)}_{\mathrm{in}} = \phi^{\mathrm{am}} + \sum_{j\,\in\,I,\,j\neq i} \phi^{(j)}_{\mathrm{sc}} \quad \text{for all}\quad i\,\in\,I.
\end{equation} 
This system of $N_f$ equations can be solved after (i) writing the truncated cylindrical eigenfunction expansion of $\phi^{\mathrm{am}}$ in the local coordinate system associated with floe $i$, (ii) applying Graf's addition theorem \cite{abramowitz_stegun70} to express $\phi^{(j)}_{\mathrm{sc}}$ in the local coordinate system associated with floe $i$ and (iii) using the exterior DTM of floe $i$ to express the expansion coefficients of $\phi^{(j)}_{\mathrm{sc}}$ in terms of those of $\phi^{(i)}_{\mathrm{in}}$. 
This procedure yields a coupled system for the coefficients $a_{m,n}^{(i)}$, which can be inverted directly \cite{kagemoto_yue86,peter_meylan04,montiel_etal13b} or solved iteratively \cite{ohkusu74,mavrakos_koumoutsakos87}.
The local scattered field coefficients $b_{m,n}^{(i)}$ and $c_{m,n}^{(i)}$ are then computed directly using \eqref{eq:DTMs}.

Numerical issues discussed by Montiel et al. \cite{montiel_etal15b,montiel_etal16} arise when the number floes becomes larger than O(100) and/or when the floes are closely spaced.
As a remedy, Montiel et al. proposed an algorithm that combines the direct approach with a domain decomposition technique, referred to as the slab-clustering method (SCM), in order to resolve wave interactions with O($10^4$--$10^5$) floes.
We only give a brief summary of the key steps of the SCM here and the reader is referred to the original papers \cite{montiel_etal15b,montiel_etal16} for further details.

\begin{enumerate}
	\item We cluster the array of floes into $N_s$ slab regions parallel to the $y$-axis, so that each slab $q$ is bounded by $x=\xi_{q-1}$ and $x=\xi_q$, with $\xi_{q-1}<\xi_q$, and contains the centre of $N^{(q)}$ floes.
	In each slab $q$, we apply the direct method summarised above to obtain the matrix mapping 
	\begin{equation}\label{eq:scat_map_polar_slab}
	\mathbf{a}_q = \mathbf{S}_q \mathbf{f}_q
	\end{equation}
	between the vectors $\mathbf{a}_q$ containing the coefficients of the locally incident field $\phi^{(i)}_{\mathrm{in}}$ on each floe $i=1,\ldots,N^{(q)}$ in slab $q$, and $\mathbf{f}_q$ containing the coefficients of the forcing field composed of the ambient field and the field scattered by adjacent slabs expressed in the local polar coordinates of each floe in slab $q$.
	
	\item We decompose the potential at each interface $x=\xi_q$, $0\le q\le N_s$, as $\phi = \phi_q^+ +  \phi_q^-$, where $\phi_q^{\pm}$ is a field propagating or decaying in the positive/negative $x$-direction. 
	Neglecting the vertical evanescent wave modes associated with the imaginary roots of \eqref{eq:DispRelOW}, we approximate these components as
	\begin{equation}\label{eq:fields_slab}
	\phi_q^{\pm}(\mathbf{x}) \approx \psi_0(z)\int_{\Lambda}A_q^{\pm}(\chi) \expo^{\cvar k_0(\pm(x-\xi_q)\cos\chi + y\sin\chi)}\ud{\chi}.
	\end{equation}
	The validity of this approximation was confirmed by Montiel et al. \cite{montiel_etal16}.
	We have introduced the unknown amplitude functions $A_q^{\pm}(\chi)$ and the directional parameter $\chi\,\in\,\Lambda$, where $\Lambda$ is the integration contour which extends into the complex plane.
	It is defined by $\Lambda=\Lambda_i^-\cup\Lambda_r\cup\Lambda_i^+$, where $\Lambda_i^{\pm} = \pm\pi/2\mp(0,\infty)$ and $\Lambda_r=[-\pi/2,\pi/2]$.
	A value of $\chi\,\in\,\Lambda_r$ corresponds to a plane wave travelling at angle $\chi$ with respect to the $x$-axis, while $\chi\,\in\,\Lambda_i^{\pm}$ corresponds to an evanescent wave decaying exponentially with $x$.
	Such evanescent wave components are generated by wave sources of the form \eqref{eq:EigFunExpSc} from floes present in the adjacent slabs.
	
	\item The amplitude functions $A_{q-1}^{\pm}(\chi)$ and $A_q^{\pm}(\chi)$, respectively defined on the left and right boundary of slab $q$, are related through the following integral scattering relationships
	\begin{subequations}\label{eq:scat_rel_slab}
		\begin{equation}\label{eq:scat_rel_slab1}
		A_{q-1}^{-}(\chi) = \int_{\Lambda}\left(\mathcal{R}_q^-(\chi:\tau)A_{q-1}^{+}(\tau) + \mathcal{T}_q^-(\chi:\tau)A_{q}^{-}(\tau)\right)\ud{\tau} \quad \text{and}
		\end{equation}
		\begin{equation}\label{eq:scat_rel_slab2}
		A_{q}^{+}(\chi) = \int_{\Lambda}\left(\mathcal{T}_q^+(\chi:\tau)A_{q-1}^{+}(\tau) + \mathcal{R}_q^+(\chi:\tau)A_{q}^{-}(\tau)\right)\ud{\tau},
		\end{equation}
		where $\mathcal{R}_q^{\pm}(\chi:\tau)$ and $\mathcal{T}_q^{\pm}(\chi:\tau)$ are the so-called reflection and transmission kernels of slab $q$.
		Semi-analytical expressions for these kernels were derived in \cite{montiel_etal16}.
		They are obtained by combining \eqref{eq:scat_map_polar_slab} with mappings between cylindrical wave fields and plane wave fields.
	\end{subequations}
	
	\item We approximate numerically the scattering relationships \eqref{eq:scat_rel_slab} by discretising the truncated integration contour  $\widetilde{\Lambda}=\widetilde{\Lambda}_i^-\cup\Lambda_r\cup\widetilde{\Lambda}_i^+$, where $\widetilde{\Lambda}_i^{\pm} = \pm\pi/2 \mp \cvar(0,\delta)$ for some $\delta\ge0$, sampling the amplitude and kernel functions at the $N_{\Lambda}$ discrete $\chi$ and $\tau$ values, and integrating this equation numerically (composite trapezoidal rule). 
	We then obtain the following matrix equations
	\begin{equation}\label{eq:scat_rel_slab_discr}
	\mathbf{A}_{q-1}^{-} = \mathbf{R}_q^-\mathbf{A}_{q-1}^{+} + \mathbf{T}_q^-\mathbf{A}_{q}^{-} 
	\quad \text{and} \quad 
	\mathbf{A}_{q}^{+} = \mathbf{T}_q^+\mathbf{A}_{q-1}^{+} + \mathbf{R}_q^+\mathbf{A}_{q}^{-},
	\end{equation}
	where $\mathbf{A}_{q}^{\pm}$ are column vectors containing the sampled values of $A_q^{\pm}(\chi)$, and $\mathbf{R}_q^{\pm}$ and $\mathbf{T}_q^{\pm}$ are square matrices containing sampled values of the reflection and transmission kernels and the quadrature weights.
	
	\item We solve the set of $2N_s$ matrix equations defined by \eqref{eq:scat_rel_slab_discr} for the amplitude vectors $\mathbf{A}_{q}^{\pm}$, $q=1,\ldots,N_s$, using the iterative $S$-matrix method of Ko and Sambles \cite{ko_sambles88}.
	This requires initialisation of the forcing amplitudes as
	\begin{equation}
	A_0^{+}(\tau) = A^{\mathrm{am}}(\tau)\expo^{\cvar k \xi_0\cos\tau} \quad \text{and} \quad A_{N_s}^{-}(\tau) = 0.
	\end{equation}
	
	\item The local scattered wave fields in each slab $q$ are obtained after transforming the plane wave forcing fields $\phi_{q-1}^{+}$ and $\phi_{q}^{-}$ into cylindrical regular wave fields with amplitudes contained in $\mathbf{f}_q$ and successively applying \eqref{eq:scat_map_polar_slab} and \eqref{eq:DTMs}.
\end{enumerate}

Convergence of the numerical method described here depends on the truncation parameters $M$ and $N$ in approximating cylindrical series expansions \eqref{eq:EigFunExp}, and $\delta$ and $N_{\Lambda}$ in approximating the plane wave expansions \eqref{eq:fields_slab}.
For the computations carried out in this study, we set $M=0$, $N=15$, $\delta=1.2$ and $N_{\Lambda}=\text{O}(100)$ (depending on the wave frequency), allowing us to compute scattered wave coefficients with three-digit accuracy.
A comprehensive convergence analysis is conducted in \cite{montiel_etal16} to justify these values.

\newpage

\section{Floe breakup criterion}\label{sec:BrkpCrit}

\subsection{Mohr-Coulomb stress}\label{sec:MC_Stress}

The thin elastic plate model of an ice floe considered here, with its underlying assumption of plane stress, allows us to write the Cauchy stress and strain tensors at each point as
\begin{equation}\label{eq:stress/strain_tensor}
\mathbf{\mathcal{S}}(r_i,\theta_i,t) = \left(
\begin{array}{cc}
\sigma_{r} & \sigma_{r\theta} \\
\sigma_{r\theta} & \sigma_{\theta}
\end{array}\right) \quad \text{and} \quad 
\mathbf{\mathcal{E}}(r_i,\theta_i,t) = \frac{D}{2}\left(
\begin{array}{cc}
\varepsilon_{r} & \varepsilon_{r\theta} \\
\varepsilon_{r\theta} & \varepsilon_{\theta}
\end{array}\right),
\end{equation}
respectively, for any floe $i\,\in\,I$. 
The tensorial components with subscript $r$ and $\theta$ are the normal stresses/strains in the radial and azimuthal direction, respectively, while the component with subscript $r\theta$ denotes the shear stress/strain.
We can express the components of the strain tensor as \cite{meylan_squire96}
\begin{equation}\label{eq:strain_tensor_comp}
	\varepsilon_{r} = \partial_{r_i}^2 w^{(i)}, \quad \varepsilon_{\theta} = \frac{1}{r_i^2}\partial_{\theta_i}^2 w^{(i)} + \frac{1}{r_i} \partial_{r_i} w^{(i)} \quad \text{and} \quad \varepsilon_{r\theta} = \frac{1}{r_i}\partial_{r\theta_i}^2 w^{(i)} - \frac{1}{r_i^2} \partial_{\theta_i} w^{(i)}.
\end{equation}
Note that a typographical error in the expression of the shear strain given in \cite{meylan_squire96} has been corrected here.
We have defined the vertical displacement $w^{(i)}\equiv w^{(i)}(r_i,\theta_i,t)$ of floe $i$, which can be related to the potential on the under surface of the floe through the kinematic condition
\begin{equation}\label{eq:kin_cond_under_floe}
w^{(i)} = \Real\left(\frac{1}{\alpha}\partial_z\phi_{\mathrm{int}}^{(i)} \expo^{- \cvar \omega t}\right) \quad \text{on} \quad z=-d \quad \text{for} \quad r_i\le a_i.
\end{equation}

At each point of this under surface and at a given time $t$, we then compute the components of the strain tensor from \eqref{eq:strain_tensor_comp} after using \eqref{eq:kin_cond_under_floe} and \eqref{eq:EigFunExpInt} to express the vertical displacement.
Note that asymptotic formulas must be used for $\varepsilon_{\theta}$ and $\varepsilon_{r\theta}$ in the limit $r_i \to 0$ (see \cite{montiel_squire15}). 

The components of the stress tensor $\mathbf{\mathcal{S}}$ are related to those of the strain tensor $\mathbf{\mathcal{E}}$ through Hooke's law
\begin{equation}\label{eq:hookes_law}
	\left(
	\begin{array}{c}
		\sigma_{r} \\
		\sigma_{\theta} \\
		\sigma_{r\theta}
	\end{array}
	\right) = \frac{ED}{2(1-\nu^2)}\left(
	\begin{array}{ccc}
		1 & \nu & 0 \\
		\nu & 1 & 0 \\
		0 & 0 & 1-\nu \\
	\end{array}
	\right) 
	\left(
	\begin{array}{c}
		\varepsilon_{r} \\
		\varepsilon_{\theta} \\
		\varepsilon_{r\theta}
	\end{array}
	\right).
\end{equation}
We can diagonalise the stress tensor as $\mathbf{\mathcal{S}} = \tilde{\mathbf{V}} \mathbf{\mathcal{D}} \tilde{\mathbf{V}}^{\mathrm{T}}$, where 
\begin{eqnarray}
\tilde{\mathbf{V}} = \left[\tilde{\mathbf{v}}_1, \tilde{\mathbf{v}}_2\right] \quad \text{and} \quad \mathbf{\mathcal{D}} = \diag\left\{\sigma_1,\sigma_2\right\}.
\end{eqnarray}
The eigenvalues $\sigma_1\equiv\sigma_1(r_i,\theta_i,t)$ and $\sigma_2\equiv\sigma_2(r_i,\theta_i,t)$ of $\mathbf{\mathcal{S}}$ are referred to as the principal stresses. 
The corresponding normalised eigenvectors $\tilde{\mathbf{v}}_1(r_i,\theta_i,t)$ and $\tilde{\mathbf{v}}_2(r_i,\theta_i,t)$ are orthogonal and define the so-called principal directions for which the shear stress vanishes in the polar coordinate frame centred at $(r_i,\theta_i)$.
In the Cartesian frame with origin at the centre of floe $i$, the matrix of eigenvectors becomes 
\begin{equation}
\mathbf{V}(r_i,\theta_i,t) = 
\left(
\begin{array}{cc}
\cos\theta_i & -\sin\theta_i \\
\sin\theta_i & \cos\theta_i
\end{array}
\right) \tilde{\mathbf{V}}(r_i,\theta_i,t).
\end{equation}

We now derive a criterion for the breakup of an ice floe which incorporates some distinctive mechanical properties of sea ice.
There exists a range of models for the mechanical failure of a large variety of materials under different types of loads (see \cite{gross_seelig11}, Chapter 2).
They take the form $F(\sigma_1,\sigma_2)\ge0$, where $F=0$ denotes a curve in the $(\sigma_1,\sigma_2)$-plane that corresponds to the onset of mechanical failure, and is often referred to as the yield curve.
Here, we associate the ``yielding'' of an ice floe with its fracture, which is reasonable for sea ice experiencing strain rates associated with wave periods of $5\text{--}20\,\text{s}$ where plastic yield is negligible \cite{cole_etal98}.

To our knowledge, no study has attempted to determine an appropriate model of wave-induced thin plate failure for sea ice, amongst the range of existing models in the literature of fracture mechanics.
For this reason, we choose the simplest model of mechanical failure, applicable to both fracture and yield, and used for materials that exhibit significantly different values of tensile and compressive strengths, defined as the maximum tensile and compression stresses that sea ice can experience before fracturing, respectively.
In the case of sea ice, tensile strength $\sigma_t$ is typically an order of magnitude lower than compression strength $\sigma_c$ \cite{timco_weeks10}.

\begin{center}
	\begin{figure}
		\includegraphics[width=1\textwidth]{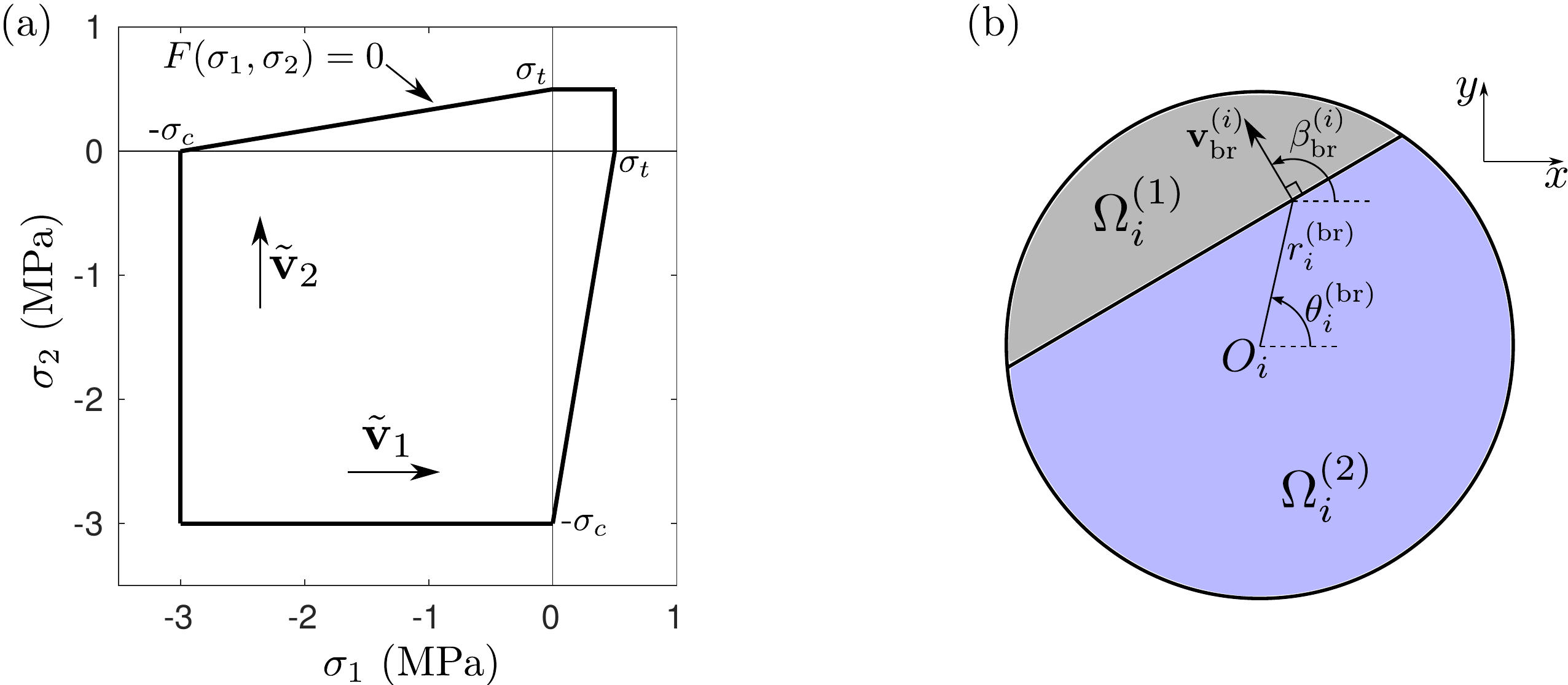}
		\caption{(a) Mohr-Coulomb yield curve $F(\sigma_1,\sigma_2)=0$ for a sea ice thin plate with tensile strength $\sigma_t=0.5\,\text{MPa}$ and compressive strength $\sigma_t=3\,\text{MPa}$. 
			The principal directions $\tilde{\mathbf{v}}_1$ and $\tilde{\mathbf{v}}_2$ are also indicated.
			(b) Partition of the surface of floe $i$ into two regions $\Omega_i^{(1)}$ and $\Omega_i^{(2)}$, resulting from the Mohr-Coulomb failure criterion \eqref{eq:brkp_crit} being satisfied.}
		\label{fig:yield_curve_brkp_ex}
	\end{figure}
\end{center}

The failure model we use here is referred to as the Mohr-Coulomb criterion \cite{gross_seelig11}.
The yield curve is defined by
\begin{equation}
F(\sigma_1,\sigma_2) \equiv \sigma^{(\mathrm{max})}(\sigma_1,\sigma_2) - \sigma^{\mathrm{MC}} = 0,
\end{equation}
where 
\begin{equation}
\sigma^{(\mathrm{max})}(\sigma_1,\sigma_2) = \max\left\{\left|\sigma_1-\sigma_2\right|+K(\sigma_1+\sigma_2), \left|\sigma_1\right|+K\sigma_1, \left|\sigma_2\right|+K\sigma_2\right\},
\end{equation}
with $K=(\sigma_c-\sigma_t)/(\sigma_c+\sigma_t)$, and
\begin{equation}
\sigma^{\mathrm{MC}}=\frac{2\sigma_c\sigma_t}{\sigma_c + \sigma_t}.
\end{equation}
We refer to $\sigma^{(\mathrm{max})}(r_i,\theta_i,t)$ as the Mohr-Coulomb stress at the point $(r_i,\theta_i)$ and time $t$, and $\sigma^{\mathrm{MC}}$ as the Mohr-Coulomb critical stress.
The yield curve $F=0$ is depicted in figure \ref{fig:yield_curve_brkp_ex}(a).
Neglecting the effect of fatigue, $F$ is assumed to be stationary.

We now define the criterion for floe breakup as follows: a floe $i\,\in\,I$ fractures into two smaller floes if the condition
\begin{equation}\label{eq:brkp_crit}
\sigma_{\mathrm{br}}^{(i)} \equiv \max\left\{\sigma^{(\mathrm{max})}(r_i,\theta_i,t), \quad \text{for} \quad (r_i,\theta_i,t)\,\in\,[0,a_i)\times[0,2\pi)\times[0,2\pi/\omega)\right\} \ge \sigma^{\mathrm{MC}}
\end{equation}
is satisfied. 
We refer to $\sigma_{\mathrm{br}}^{(i)}$ as the potential breakup stress of floe $i$.
The breakup criterion depends on the tensile and compressive strength of sea ice, which we estimate to be $\sigma_c=3\,\text{MPa}$ and $\sigma_t=0.5\,\text{MPa}$, respectively.
These values were chosen from empirical formulas reported by Timco and Weeks \cite{timco_weeks10}, assuming a brine volume fraction of approximately 0.04 and a strain rate of $2\times10^{-5}\,\text{s}^{-1}$ consistent with a loading at a mid-range wave period of 10\,s.
It should be noted that flexural strength of sea ice may be a better approximation for $\sigma_t$ than the tensile strength used here for flexurally-induced fracture. 
The value calculated from the empirical formula given in \cite{timco_weeks10}, however, is close to that for the tensile strength, so the distinction is actually of no practical importance.

\subsection{Potential breakup stress in a single floe}
\label{sec:BrkpStress_1floe}

We devise a sensitivity test to assess the potential for breakup of a single ice floe (i.e.\ $N_s=1$ and $N^{(1)}=1$) for a range of radii and wave periods. 
We prescribe a unidirectional plane wave with unit amplitude travelling in the positive $x$ direction, by setting $A^{\textrm{am}}(\tau) = \delta(\tau)$, where $\delta$ denotes the Dirac delta.
We fix the water depth to $h=200\,\text{m}$.
We assume, without loss of generality, that the floe has its centre coinciding with the origin $(x,y)=(0,0)$ of the horizontal Cartesian coordinate system. 
We compute the potential breakup stress $\sigma_{\mathrm{br}}$ for wave periods $T=5\text{--}20\,\text{s}$, floe radii $a=5\text{--}500\,\text{m}$ and floe thicknesses $D=1$, 2 and 4\,m.

\begin{figure}
	\begin{center}
		\includegraphics[width=1\textwidth]{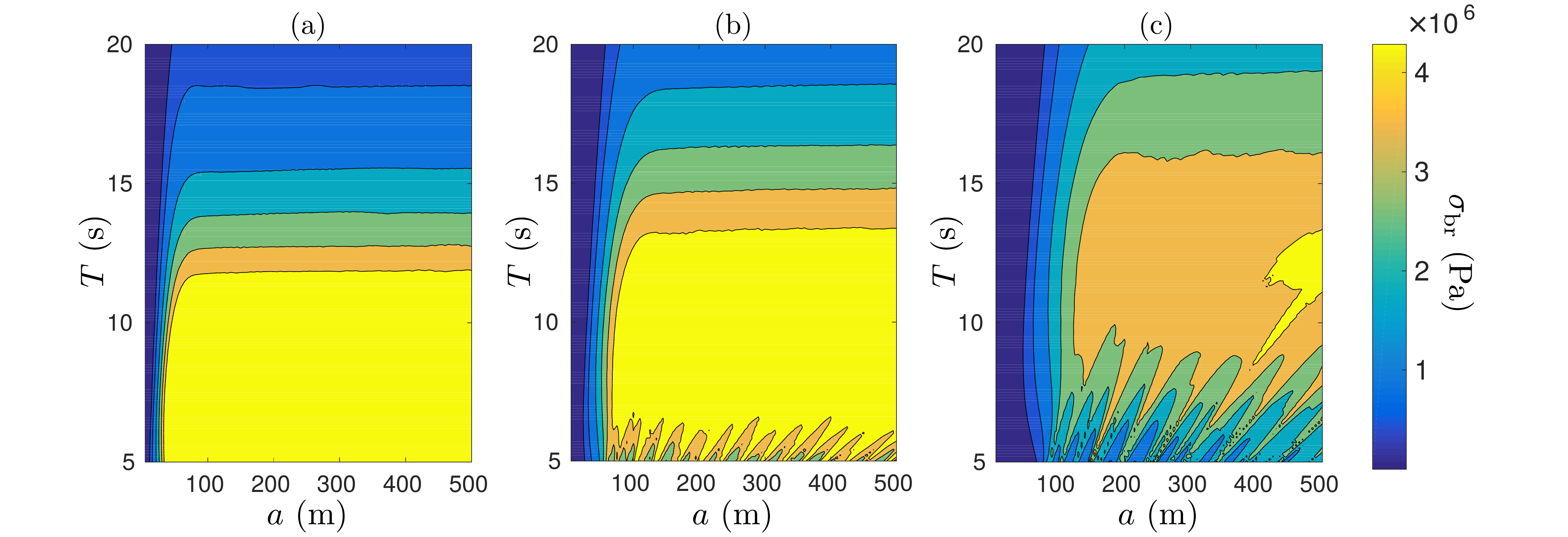}
		\caption{Filled contour plots of the potential breakup stress $\sigma_{\mathrm{br}}$ against wave period $T$ and floe radius $a$, under unit amplitude plane wave forcing and for a single floe of thickness (a) $D=1\,\text{m}$, (b) $D=2\,\text{m}$ and (c) $D=4\,\text{m}$.}
		\label{fig:MCStress}
	\end{center}
\end{figure}

Filled contour plots of $\sigma_{\mathrm{br}}$ against wave periods and floe radii are shown in figure \ref{fig:MCStress}(a), (b) and (c) for $D=1$, 2 and $4\,\text{m}$, respectively.
The six contours displayed in each plot correspond to values of the potential break stress $\sigma_{\mathrm{br}}=e\sigma^{\mathrm{MC}}$ for $e=0.5, 1, 2, 3, 4$ and $5$, noting that $\sigma^{\mathrm{MC}}\approx0.86\,\text{MPa}$.

For all three thicknesses considered, we generally observe that the potential breakup stress increases rapidly with the floe size before plateauing for floe radii greater than a certain value.
This behaviour is seen at all wave periods for $D=1\,\text{m}$ and for periods greater than approximately 6 and 9\,s for $D=2$ and 4\,m, respectively.
For shorter waves, $\sigma_{\mathrm{br}}$ oscillates with respect to $a$, suggesting resonances periodically induce large stress values, as the floe size approximately equals an integer multiple of the wavelength. 

\begin{figure}
	\begin{center}
		\includegraphics[width=1\textwidth]{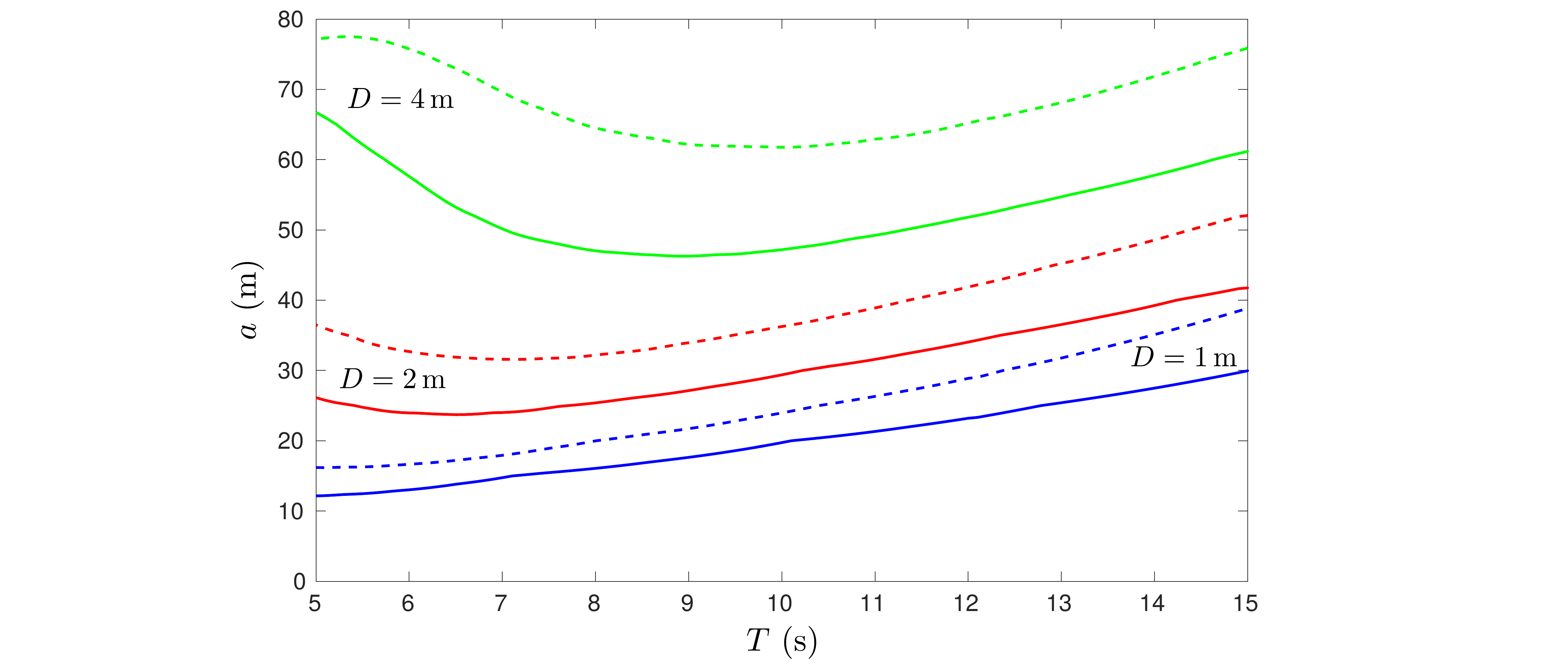}
		\caption{Variation of the Mohr-Coulomb radius $a^{\mathrm{MC}}$ (dashed lines) and the critical breakup radius $a_{\mathrm{crit}}$ (solid lines) with respect to $T$ for floe thicknesses $D=1\,\text{m}$ (blue lines), $D=2\,\text{m}$ (red lines) and $D=4\,\text{m}$ (green lines).}
		\label{fig:brkp_contour}
	\end{center}
\end{figure}

According to our breakup criterion \eqref{eq:brkp_crit}, fracture occurs for $e\ge1$, corresponding to the second contour in figure \ref{fig:MCStress}.
For all wave periods, it is seen that breakup occurs for all floe radii larger than a critical radius denoted by $a^{\mathrm{MC}}$ and referred to as the Mohr-coulomb radius. 
It is shown as a function of wave period in figure \ref{fig:brkp_contour} for the three ice thicknesses considered (dashed lines).
We observe that it reaches a minimum at $T=7$ and 10\,s for $D=2$ and 4\,m, respectively, while it seems to approach a minimum near $T=5\,\text{s}$ for $D=1\,\text{m}$.
These correspond to resonant frequencies, when the floe diameter approximately coincides with the open water wavelength and half the ice-covered wavelength.

In figure \ref{fig:brkp_contour}, we further indicate by solid lines the radius corresponding to the contour defined by $e=0.5$ in figure \ref{fig:MCStress}.
We denote it by $a_{\mathrm{crit}}$ and refer to it as the critical breakup radius.
Although breakup does not occur for these radii in the single floe scattering simulations conducted here with a unit amplitude plane incident wave, we envisage that multiple interacting floes may generate more energetic wave forcing with the ability to fracture smaller floes as a result of constructive interference.
In all subsequent simulations, we set $a_{\mathrm{crit}} = a_{\mathrm{crit}}(T)$ as the minimum floe radius below which breakup cannot occur, unless otherwise discussed.
It should be noted that our arbitrary choice of $a_{\mathrm{crit}}$ is conservative in most cases in the sense that the probability that a floe with $a<a_{\mathrm{crit}}$ will fracture is small compared to the probability that a floe with $a>a_{\mathrm{crit}}$ does not.

\section{Breakup model}\label{sec:BrkpModel}

We now seek to model the breakup of an array of ice floes, with the goal of determining the evolution of the floe size distribution (FSD) towards a steady state under repeated wave action and breakup events.
To do this, we propose a numerical procedure that simulates the repeated breakup of an ice cover initially composed of $N_f$ identical large floes with radius $a_{\mathrm{max}}$.
To reduce the number of single floe solutions that needs to be computed, we consider a finite number $N_r$ of floe radii $a^{(1)},\ldots,a^{(N_r)}$, such that $a^{(1)}=5\,\text{m}$ and $a^{(N_r)}=a_{\mathrm{max}}$.
We create a row vector $\mathrm{V}^{(\mathrm{FSD})}_0$ of length $N_r$, such that the $l$th entry contains the number of floes in the array with radius $a^{(l)}$, for $l=1,\ldots,N_r$.
For the initial configuration of floes, we then have
\begin{equation}
\mathrm{V}^{(\mathrm{FSD})}_0 = \left(0,\ldots,0,N_f\right).
\end{equation}

The algorithm used for our breakup simulations is outlined as follows below.
\begin{enumerate}
	\item Compute the exterior and interior DTMs associated with each floe radius $a^{(p)}$, for $p=1,\ldots,N_r$, using the method discussed in \textsection\ref{sec:ScatModel}\ref{sec:SingleFloeScat} and store them (the size of each DTM is O(10)).
	
	\item Compute the solution of the multiple scattering problem using the method presented in \textsection\ref{sec:ScatModel}\ref{sec:MultiFloeScat} for the initial configuration of ice floes.
	
	\item Compute the potential breakup stress $\sigma_{\mathrm{br}}^{(i)}$ for each floe $i\,\in\,I$ directly from the definition given in \eqref{eq:brkp_crit}.
	In addition, define the normal breakup direction $\mathbf{v}_{\mathrm{br}}^{(i)}$ as the vector normal to the yield curve in the local Cartesian coordinate system of floe $i$, that is
	\begin{equation}
	\mathbf{v}_{\mathrm{br}}^{(i)} = \left(
	\begin{array}{c}
	\cos\beta_{\mathrm{br}}^{(i)} \\
	\sin\beta_{\mathrm{br}}^{(i)}
	\end{array}\right) = \mathbf{V}(r_i^{(\mathrm{br})},\theta_i^{(\mathrm{br})},t_i^{(\mathrm{br})})\frac{\nabla_{\sigma}F}{\left|\nabla_{\sigma}F\right|}
	\end{equation}
	where $t_i^{(\mathrm{br})}$ and $(r_i^{(\mathrm{br})},\theta_i^{(\mathrm{br})})$ are the time and polar coordinates of the point of floe $i$, respectively, at which $\sigma_{\mathrm{br}}^{(i)}$ is computed, and $\nabla_{\sigma}=(\partial_{\sigma_1},\partial_{\sigma_2})^{\mathrm{T}}$.
	
	\item For each floe $i$, test the breakup criterion \eqref{eq:brkp_crit}. 
	If the inequality is not satisfied, floe $i$ does not fracture and therefore remains in the array.
	If the inequality holds, however, remove floe $i$ with radius $a_i$ from the array and substitute it with two floes of radii $a_i^{(1)}<a_i$ and $a_i^{(2)}<a_i$ defined as follows:
	at the point $(r_i^{(\mathrm{br})},\theta_i^{(\mathrm{br})})$, draw a straight line perpendicular to the vector $\mathbf{v}_{\mathrm{br}}^{(i)}$, partitioning the region $\Omega_i$ into two regions $\Omega_i^{(1)}$ and $\Omega_i^{(2)}$ as shown in figure \ref{fig:yield_curve_brkp_ex}(b). 
	We then define $a_i^{(1)}$ and $a_i^{(2)}$ as the radii of the disks with the area of regions $\Omega_i^{(1)}$ and $\Omega_i^{(2)}$, respectively.
	Their expressions are
	\begin{equation}
	a_i^{(1)} = a_i\sqrt{\frac{\pi - 2\theta_0 - \sin2\theta_0}{2\pi}} \quad \text{and} \quad a_i^{(2)} = a_i\sqrt{\frac{\pi + 2\theta_0 + \sin2\theta_0}{2\pi}}
	\end{equation}
	where 
	\begin{equation}
	\theta_0=\left|\sin^{-1}\left(\frac{r_i^{(\mathrm{br})}}{a_i} \cos\left(\theta_i^{(\mathrm{br})} - \beta_{\mathrm{br}}^{(i)}\right)\right)\right|.
	\end{equation}
	Note that if either $a_i^{(1)}$ or $a_i^{(2)}$ is not equal to one of the radii $a^{(l)}$, $l=1,\ldots,N_r$, we round it to the nearest one.
	
	\item Update the FSD by defining the vector $\mathrm{V}^{(\mathrm{FSD})}_1$ containing the number of floes in the new array, i.e.\ after breakup has occurred, with each radius $a^{(l)}$, $l=1,\ldots,N_r$.
	
	\item Generate a random array of circular floes described by the FSD vector $\mathrm{V}^{(\mathrm{FSD})}_1$. 
	For this purpose, we use the random array generator devised by Montiel et al.\ \cite{montiel_etal16} (see Appendix B therein).
	
	\item Repeat steps (ii)--(vi) $N_{\mathrm{br}}-1$ times, where $N_{\mathrm{br}}$ is the number of breakup events considered for the simulation. 
	At the end of each iteration $s$, we obtain an updated FSD defined by the vector $\mathrm{V}^{(\mathrm{FSD})}_s$, for $s=1,\ldots,N_{\mathrm{br}}$.
\end{enumerate}

The breakup model described here should be seen as a new method to generate an FSD from repeated wave-induced floe breakup events, as opposed to an attempt to approximate the breakup process in the MIZ.
From this perspective, the gross approximation of generating two circular floes from the breakup of a circular floe is acceptable, as we are only interested in the size of newly created floes.
This approach is analogous to that of \cite{toyota_etal11}, who measured floe size in the MIZ by calculating the diameter of disks with the same area as that of the observed floes.

In step (v), we only include floes with radius $a^{(l)} \ge a_{\mathrm{crit}}$ in $\mathrm{V}^{(\mathrm{FSD})}_s$ to generate the updated random array.
Neglecting the influence of smaller floes on the breakup simulations improves the efficiency of the scattering computation in step (ii), noting that these small floes are still counted in the vector $\mathrm{V}^{(\mathrm{FSD})}_{s+1}$.

It should be further noted that although the breakup algorithm described here is intended to preserve the ice concentration (defined as the fraction of ice-covered ocean surface), rounding the radius of the two floes generated in step (iv) introduces a small change of concentration.
However, simulations (not shown here) have revealed that these changes average out over a large number of breakup events, so the concentration actually remains quasi-constant.

\section{Results}\label{sec:Results}

\subsection{Single floe}\label{sec:ResOneFloe}

We first conduct numerical experiments with the goal of understanding the FSD generated from the breakup of a single large ice floe under monochromatic and unidirectional wave forcing, as considered in \textsection\ref{sec:BrkpCrit}\ref{sec:BrkpStress_1floe}. 
We set $N_f=1$ with $a_{\mathrm{max}}=200\,\text{m}$. 
A sensitivity study, not shown here, demonstrated that choosing $a_{\mathrm{max}}=500\,\text{m}$ resulted in similar post-breakup FSDs, so the smaller radius was chosen for computational efficiency.
We set $N_r=74$, 84 and 100 unique floe radii between 5\,m and $a_{\mathrm{max}}$ for $D=1$, 2 and $4\,\text{m}$, respectively, so that we have a 5\,m resolution for $a\ge50$, 60 and $80\,\text{m}$, respectively, and a 1\,m resolution for smaller radii.
Sensitivity tests (not shown here) indicated that the critical radius $a_{\mathrm{crit}}$, below which scattering and breakup is not likely to occur, needs to be lowered from the values discussed in \textsection\ref{sec:BrkpCrit}\ref{sec:BrkpStress_1floe} for the cases $D=4\,\text{m}$ and $T<10\,\text{s}$, probably because of the strong effect of multiple scattering.
We therefore chose $a_{\mathrm{crit}}=40$, 40, 35, 30 and 30\,m for $T=5$, 6, 7, 8 and 9\,s.

The floe breakup algorithm described in \textsection\,\ref{sec:BrkpModel} is then used to simulate the evolution of the FSD for $N_{\mathrm{br}}=50$ breakup events, which we find is generally sufficient to reach a steady state.
The ice concentration is set to 50\%, so that the random array of floes generated after each breakup event is enclosed in a square region with side length $\sqrt{2\pi}a_{\mathrm{max}}$.

The outputs of the breakup algorithm are the vectors $\mathrm{V}^{(\mathrm{FSD})}_s$, for $s=0,\ldots,50$, describing the FSD after $s$ breakup events.
These vectors can interpreted as discrete functions of the floe radius variable $a$ taking $N_r$ values. 
The floe size probability density function (PDF), denoted by $\mathcal{P}(a)$, after $s$ breakup events is defined as the linearly interpolated discrete function with values given in $\mathrm{V}^{(\mathrm{FSD})}_s$ divided by the area under its curve.
The PDFs are further averaged over 10 random realisations of the breakup simulation.

\begin{figure}
	\begin{center}
		\includegraphics[width=1\textwidth]{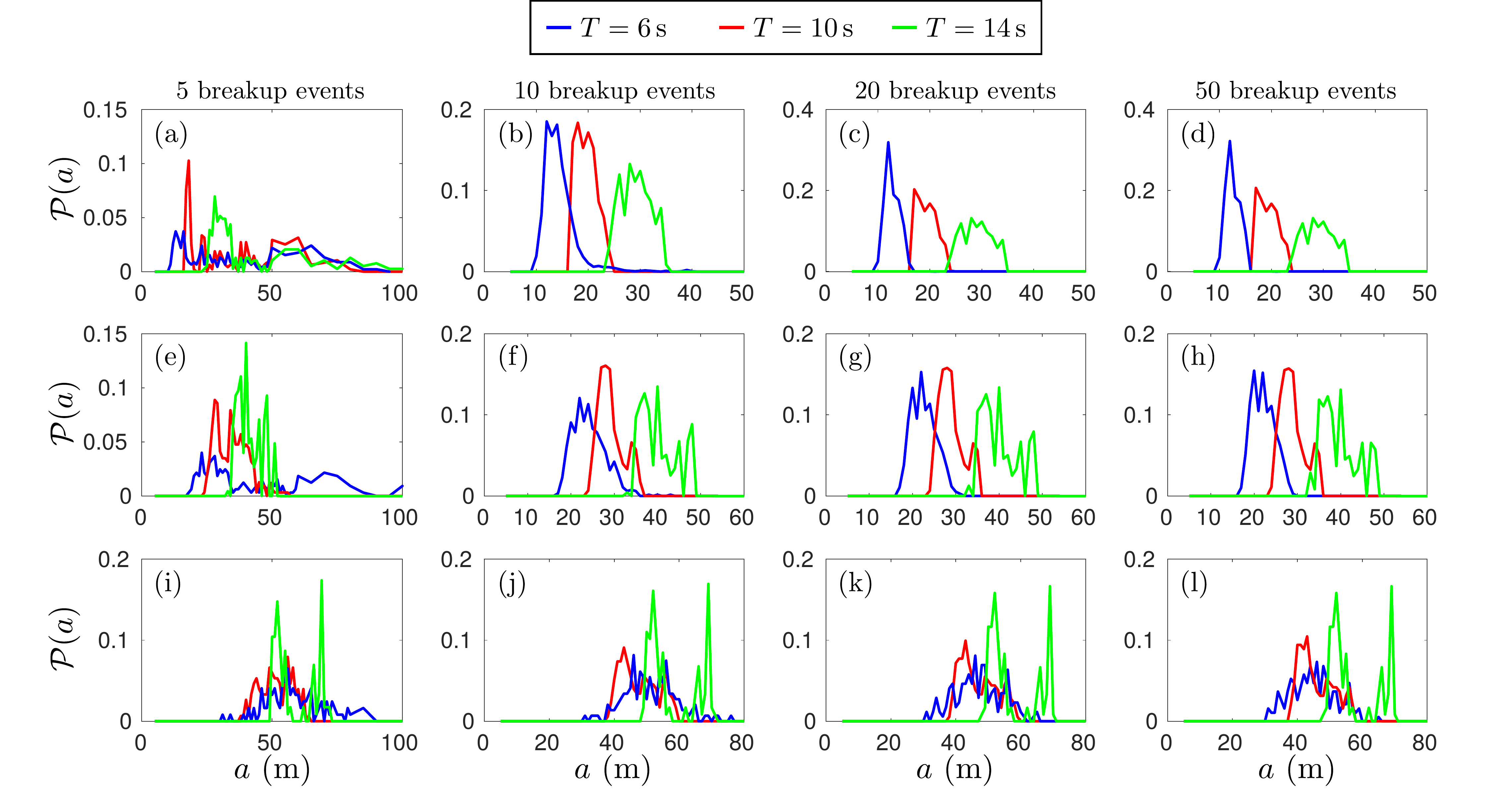}
		\caption{Evolution of the floe size PDF $\mathcal{P}(a)$ during the repeated breakup of a single floe with radius $a_{\mathrm{max}}=200\,\text{m}$. 
		Results are displayed for (a--d) $D=1\,\text{m}$, (e--h) $D=2\,\text{m}$ and (i--l) $D=4\,\text{m}$ and wave periods $T=6$, 10 and 14\,s (blue, red and green lines, respectively).
	    The PDFs are shown after (a,e,i) $s=5$, (b,f,j) 10, (c,g,k) 20 and (d,h,l) 50 breakup events.}
		\label{fig:FSD_1floe}
	\end{center}
\end{figure}

In figure \ref{fig:FSD_1floe}, the floe size PDF is plotted after $s=5$, $10$, $20$ and $50$ breakup events for the wave periods $T=6\,\text{s}$ (blue lines), $10\,\text{s}$ (red lines) and $13\,\text{s}$ (green lines) and the three ice thicknesses $D=1\,\text{m}$ (top panels), $2\,\text{m}$ (middle panels) and $4\,\text{m}$ (bottom panels). 
The first striking feature is that the distributions obtained from the breakup simulations clearly cannot be identified as power law distributions, as the number of floes decreases to zero for smaller and smaller radii.
Instead, the distributions after 50 breakup events look either uni-modal or multi-modal, as the distributions contain one or multiple maxima.
Closely spaced successive maxima, such as the ones seen for $T=6\,\text{s}$ and $D=2\,\text{m}$ around $a=20\,\text{m}$ are likely to be an artefact of the floe radius sampling chosen for the simulations, so that they actually represent a single mode of the corresponding continuous distribution.
Accounting for this, we propose that all the PDFs are either uni-modal or bi-modal.
Uni-modality is observed for $D=1\,\text{m}$ at all wave periods and $D=2\,\text{m}$ for $T=6\,\text{s}$, while bi-modality manifests itself in all other cases.
It should be further noted the uni-modal distributions are all positively skewed, suggesting they may be the superposition of two closely spaced uni-modal distributions.
The bi-modality may be explained from the fact that each floe with sufficient stress breaks into two floes only, so that the repeated breakup of the ice cover results in two dominant floe sizes that correspond approximately to the breakup of the smallest floe that can fracture for a given thickness and wave period. 
We do not attempt to analyse the bi-modal property of the PDFs further.

For each parameter configuration considered in figure \ref{fig:FSD_1floe}, we also observe a convergence of the PDFs with respect to the number of breakup events, as all distributions obtained for 20 events are almost identical to those obtained after 50 events. 
The convergence seems to occur faster for longer waves and thicker floes, which is a consequence of less floe breakups occurring for increasing values of these parameters.

\begin{figure}
	\begin{center}
		\includegraphics[width=1\textwidth]{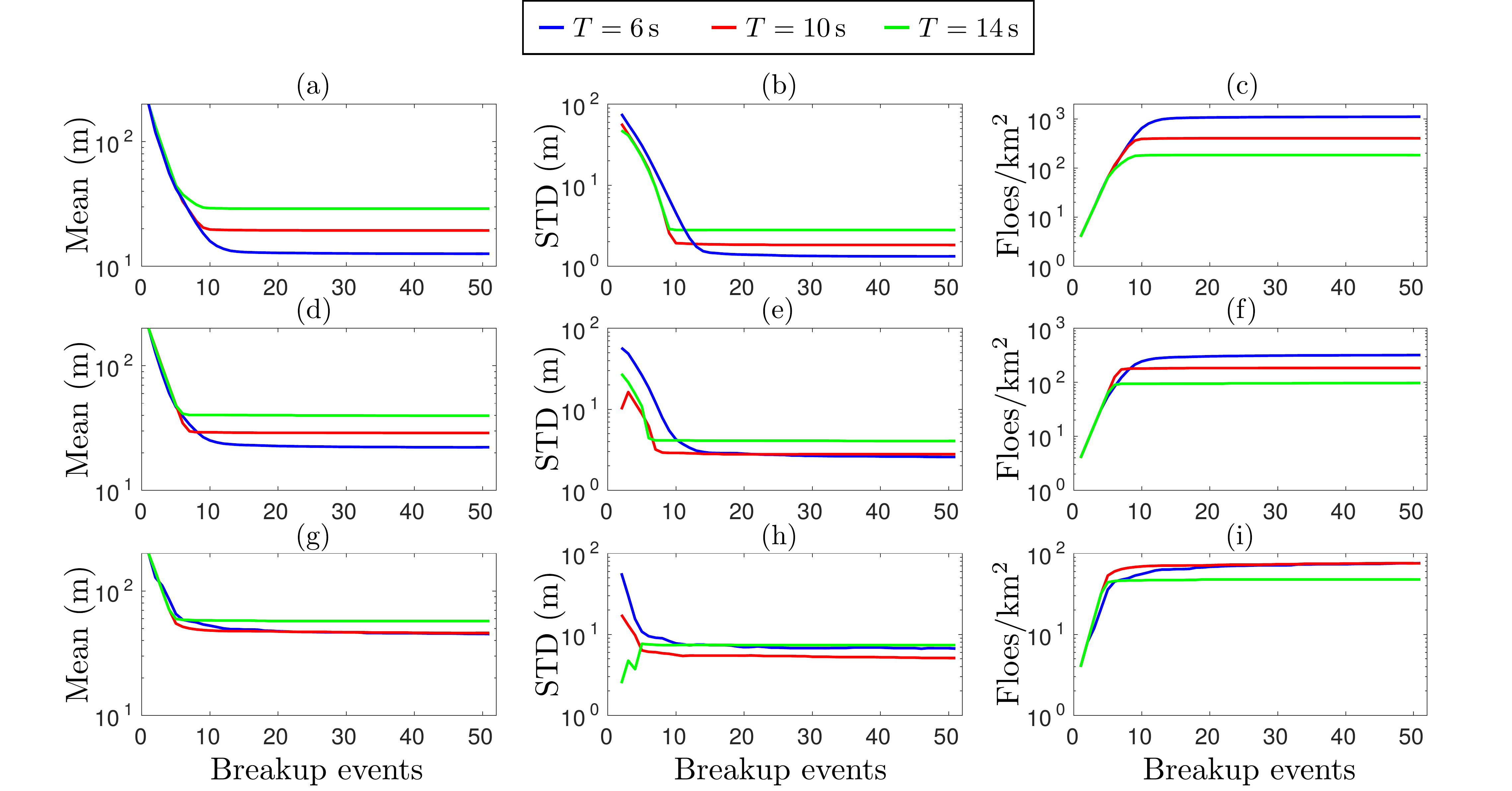}
		\caption{Evolution of (a,d,g) the mean and (b,e,h) standard deviation (STD) of the FSD, and (c,f,i) the number of floes per square kilometre, through the 50 breakup events.
		Each row of panels and line colour corresponds to a floe thickness and wave period, respectively, as defined in figure \ref{fig:FSD_1floe}.}
		\label{fig:Moments_1floe}
	\end{center}
\end{figure}

To understand better the convergence of the FSDs, figure \ref{fig:Moments_1floe} shows the evolution of the mean and standard deviation (STD) of the PDFs, and number of floes per square kilometre through the 50 breakup events, for each wave period and thickness considered in figure \ref{fig:FSD_1floe}.
In the first few breakup events (i.e.\ up to $s\approx5$), we observe that the statistics of the distribution change exponentially fast, corresponding to the regime in which all the floes in the array fracture, so that the number of floes doubles after each events.
Interestingly, the mean floe size decreases independently from wave period and floe thickness in this regime, while the standard deviation consistently remains higher at $T=6\,\text{s}$ than at the other periods for the three floe thicknesses considered, suggesting that shorter waves generate a broader FSD under intense wave-induced breakup.

For $s>5$, the statistics of the FSD quickly reach a steady state regime.
The transition between the exponential breakup and steady state regimes is very sharp for $T=14\,\text{s}$, i.e.\ within 5 more breakup events, while it is longer for shorter waves, e.g.\ at $T=6\,\text{s}$ for which it can take more than 10--15 additional events. 
In particular, for a thickness $D=4\,\text{m}$, a small number of floes are still breaking at the end of the simulation, i.e.\ $s=50$.
Two processes are hypothesised to be responsible for the extended transition regime, viz.\ (i) multiple wave scattering within the array of floes, which is expected to be strong for these parameters and to cause constructive interference that favours floe breakup, and (ii) mixing of the floes as a result of the randomisation of the array of floes after each breakup event. (See step (vi) of the breakup algorithm described in \textsection\,\ref{sec:BrkpModel}). 
Although this latter effect is an artefact of our breakup model which may amplify breakup, as large floes will ultimately end up in front of the array and then fracture, its influence on the final steady state statistics of the FSD appears to be small for the single floe breakup simulations conducted in this section.

\begin{figure}
	\begin{center}
		\includegraphics[width=1\textwidth]{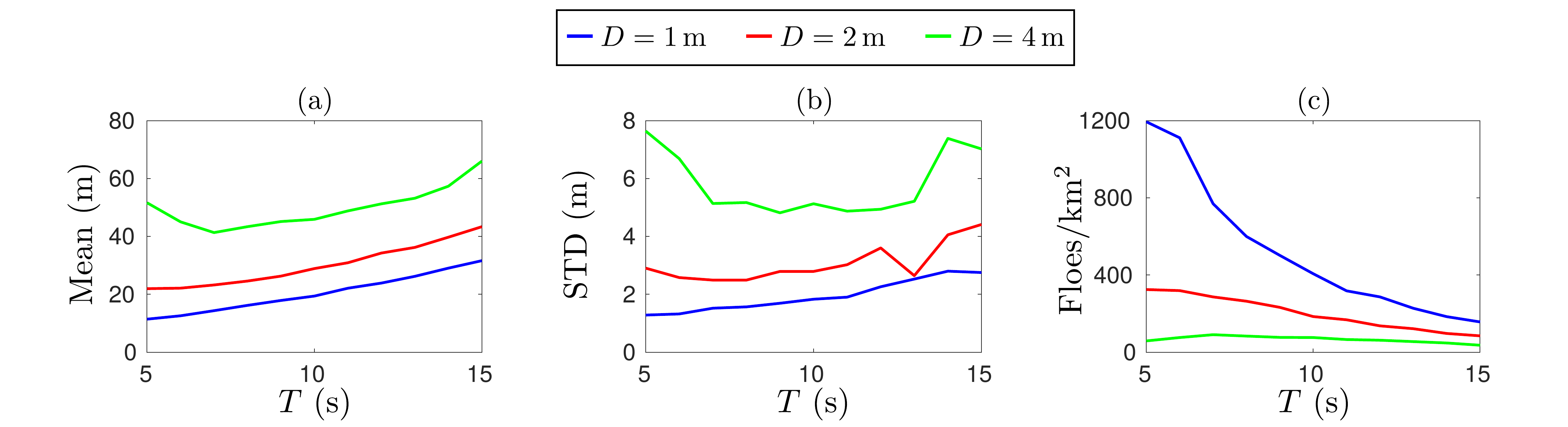}
		\caption{Steady state statistics (i.e.\ after $s=50$ breakup events) plotted against wave period for floe thicknesses $D=1$, 2 and 4\,m (blue, red and green line, respectively). 
		The statistics considered here are (a) the mean and (b) standard deviation of the FSD, and (c) the number of floes per square kilometre.}
		\label{fig:Multifreq_1floe}
	\end{center}
\end{figure}

The dependence of the steady state statistics (i.e.\ after $s=50$ breakup events) on the wave period and floe thickness is shown in figure \ref{fig:Multifreq_1floe}.
Mean and standard deviation are larger for increasing floe thicknesses, which is a result of thicker floes breaking less, as indicated in figure \ref{fig:Multifreq_1floe}(c), generating a FSD composed of a smaller number of larger floes.
The larger standard deviation is explained by the stronger bi-modality observed in the PDFs of the distribution for thicker floes, as can be seen in figure \ref{fig:FSD_1floe}, because a larger separation of the two peaks results in a wider distribution.

For $D=1$ and 2\,m, mean and standard deviation smoothly vary with wave period with a general increasing trend for longer waves.
This can also be explained by the fact that floe breakup diminishes for longer waves, which tends to enhance the formation of bi-modal distributions.
Note the minimum reached by the standard deviation at $T=13\,\text{s}$ for $D=2\,\text{m}$.
Inspection of the PDF obtained for this parameter configuration (not displayed here) shows that the bi-modal shape of the distribution is not apparent, in contrast to $T=12$ and 14\,s for which it clearly exists. 
We could not further explain this feature.

For $D=4\,\text{m}$, the mean floe size reaches a minimum at $T=7\,\text{s}$. 
It should be noted that this is smaller than $T\approx9\,\text{s}$ for which the minimum of the critical radius $a_{\mathrm{crit}}(T)$ discussed in figure \ref{fig:brkp_contour} occurs for this thickness. 
This is likely to be a consequence of multiple wave scattering enhancing the stress field in the floes within the array at smaller wave periods, and breakup ensuing due to constructive interference.
This further explains the need to take values of $a_{\mathrm{crit}}$ for the breakup simulations smaller than those computed in \textsection\ref{sec:BrkpCrit}\ref{sec:BrkpStress_1floe}, as discussed at the beginning of this section.

\subsection{Array of floes}\label{sec:ResMIZ}

We now investigate how wave scattering by an array of floes influence the breakup process and associated evolution of its FSD.
We consider the following initial configurations: (i) $N_s=10$ slabs containing 3 floes each (i.e.\ $N_f=30$) with thickness $D=1\,\text{m}$ and (ii) $N_s=20$ slabs containing 5 floes each (i.e.\ $N_f=100$) with thicknesses $D=2$ and 4\,m.
We chose a smaller array for $D=1\,\text{m}$ due to numerical constraints (thinner ice floes break up more, resulting in an increased computational cost required to solve the corresponding wave interaction problem with many floes).
All the floes have initial radius $a_{\mathrm{max}}=200\,\text{m}$ and the ice concentration is 50\%, so that arrays in configurations (i) and (ii) have approximate horizontal extent $1.5\,\text{km}\times5\,\text{km}$ and $2.5\,\text{km}\times10\,\text{km}$, respectively.
The unique floe radii and critical radii $a_{\mathrm{crit}}$ for each thickness are the same as those chosen for the single floe breakup simulations in \textsection\ref{sec:Results}\ref{sec:ResOneFloe}.
We perform the array breakup simulations for $N_{\mathrm{br}}=50$ breakup events.
As a result of the significantly higher computational cost associated with the array breakup simulations compared to the single floe breakup simulations, no ensemble averaging was performed here, so all the results in this sub-section are obtained from a single random realisation for each wave period and floe thickness.
Additional random realisations performed on a few selected cases showed remarkable consistency of the resulting FSD, suggesting very little variability exists in the stochastic process described here.

\begin{figure}
	\begin{center}
		\includegraphics[width=1\textwidth]{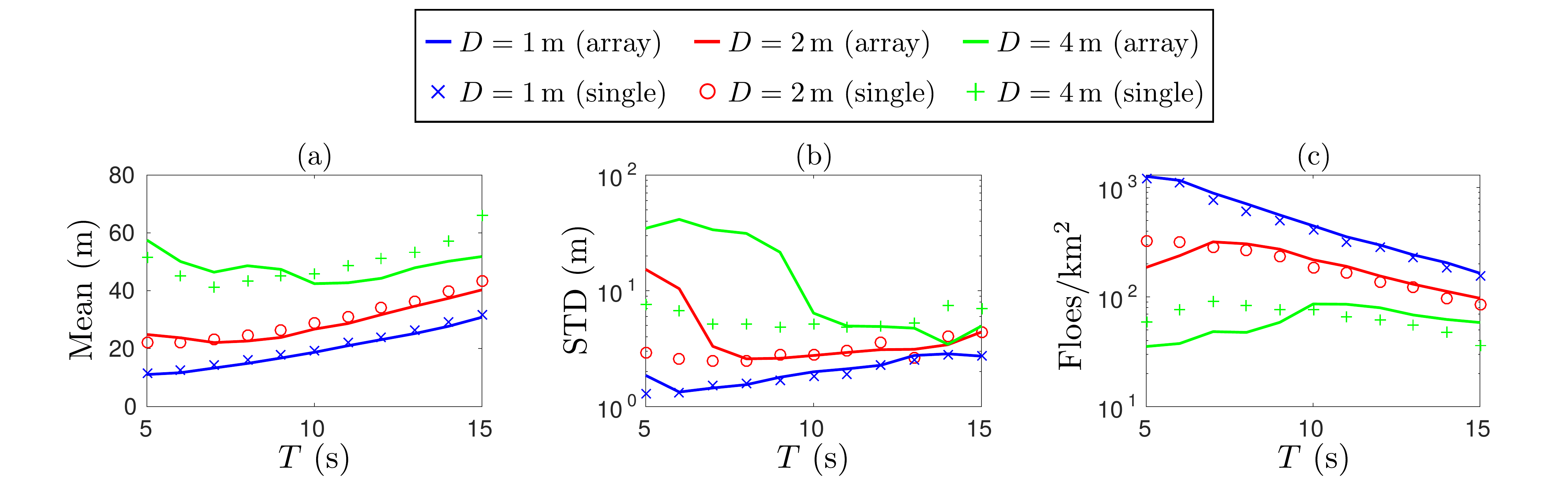}
		\caption{Same as figure \ref{fig:Multifreq_1floe} for the array simulations.}
		\label{fig:Multifreq_MIZ}
	\end{center}
\end{figure}

The mean, standard deviation and number of floes per square kilometre obtained after 50 breakup events for each wave period and floe thickness considered are plotted in figure \ref{fig:Multifreq_MIZ}, where they are compared to the steady state statistics of the corresponding single floe breakup simulations.
For $D=1\,\text{m}$, the steady-state statistics obtained by breaking up the $10\times3$ array are very similar to those obtained from the breakup of a single floe, over the range of wave periods.
It should be noted that the mean is consistently slightly smaller while the number of floes is slightly larger, suggesting more breakup takes place for the array simulations, probably as a result of enhanced floe breakup due to constructive interference caused by multiple scattering.
A similar observation can be made for $D=2\,\text{m}$-thick ice in the range of wave periods $T\ge7\,\text{s}$.
For shorter waves, there is a clear deviation from the single floe breakup statistics, as the mean and standard deviation increase and the number of floes decreases as $T$ decreases, which is the opposite trend to what happens for the single floe breakup.
This indicates that scattering is sufficiently dominant in this regime to prevent some floes from fracturing.
The large increase of the standard deviation further suggests that the FSD spread over a larger range of floe radii.
Results obtained for $D=4\,\text{m}$ reinforce our explanations of the role of multiple scattering on breakup through the large array, with the observations of two regimes, i.e.\ $T<10\,\text{s}$ (short waves) and $T\ge10\,\text{s}$ (long waves).
In the short wave regime, scattering generates sufficient wave energy attenuation to prevent floe breakup at some level of penetration in the ice-covered domain.
In the long wave regime, scattering enhances floe breakup due to constructive interference of the wave fields radiated by the freely floating individual ice floes.

\begin{figure}
	\begin{center}
		\includegraphics[width=1\textwidth]{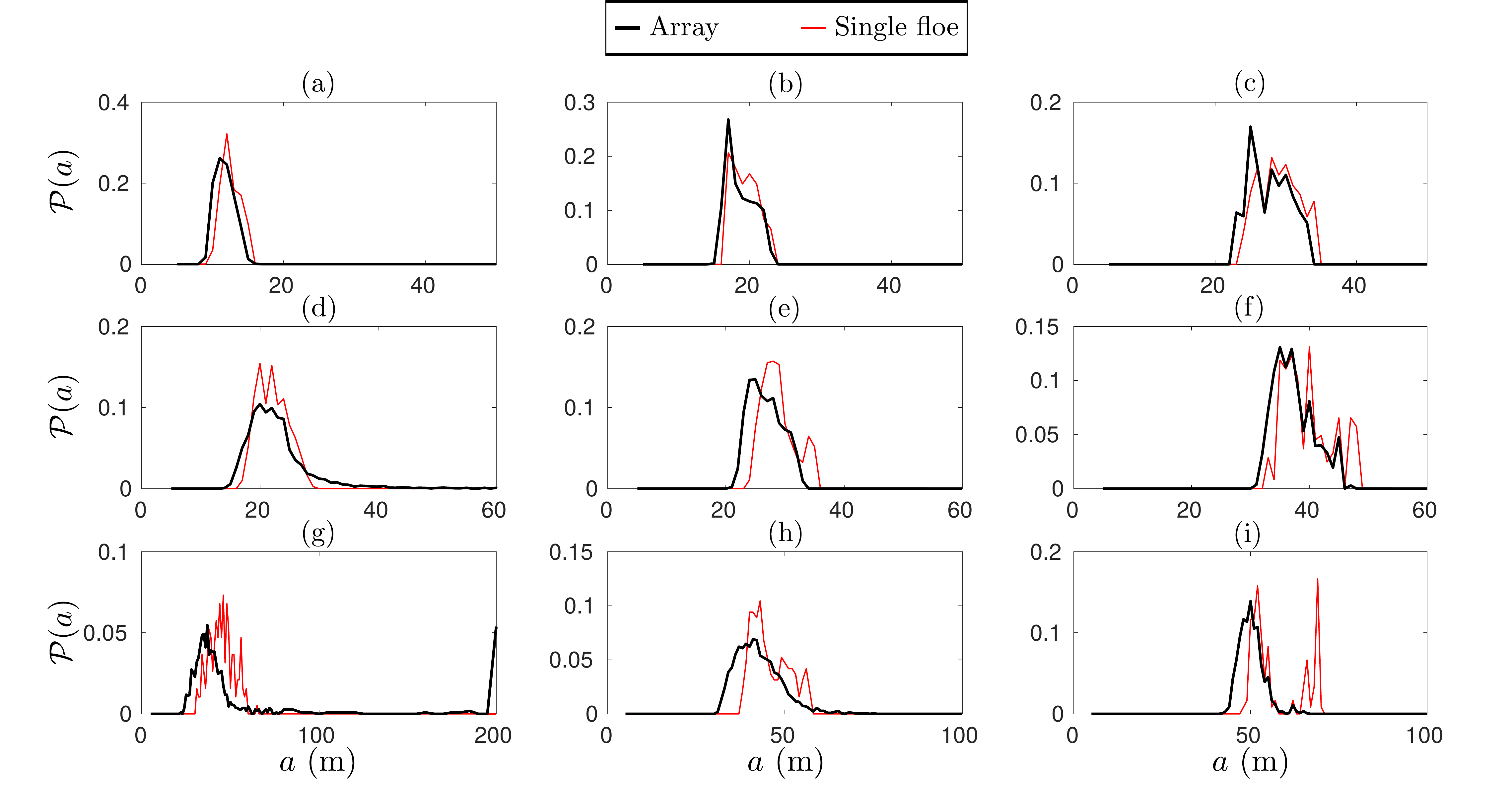}
		\caption{Comparison of the steady-state PDFs $\mathcal{P}(a)$ obtained from the breakup of a large array (thick black line) and a single floe (thin red line) at wave periods (a,d,g) $T=6\,\text{s}$, (b,e,h) $T=10\,\text{s}$ and (c,f,i) $T=14\,\text{s}$.
		Results are displayed for (a--c) $D=1\,\text{m}$, (d--f) $D=2\,\text{m}$ and (g--i) $D=4\,\text{m}$.}
		\label{fig:StSt_PDFs_MIZ}
	\end{center}
\end{figure}

We seek more insight about the FSD obtained from the breakup of large arrays by plotting the PDFs of the distributions after 50 breakup events in figure \ref{fig:StSt_PDFs_MIZ}.
We can identify the two regimes discussed previously, i.e.\ enhanced breakup and reduced breakup, as a result multiple wave scattering by a large array of floes.
We observe enhanced breakup in panels (a--c), (e,f) and (i), corresponding to cases for which scattering is not significant (i.e.\ long waves and/or thin ice).
The PDFs show the presence of a larger number of small floe compared to the single floe breakup simulations, while the presence of larger floes decreases.
Although this can be viewed as a shift of the PDF towards lower floe radii, the shape of the distribution also changes.
In particular, the bi-modality observed in the distributions obtained from the breakup of a single floe is not a persistent feature of the PDFs associated with the breakup of an array, as the second mode (i.e.\ the mode corresponding to larger floes) is damped or removed.

The second regime, characterised by reduced breakup in the array, corresponds to the PDFs shown in panels (d) and (g,h).
The large spread of these distributions discussed above is clearly observed, with an increased presence of smaller floes (compared to single floe breakup), as is the case in the long waves/thin ice regime, as well as larger floes. 
Important wave scattering occurring in the front slabs of the array prevents wave energy to cause breakup deeper in the array.
In the extreme case $D=4\,\text{s}$ and $T=6\,\text{s}$ depicted in panel (g), we can see the presence of floes with radius $a=200\,\text{m}$, which is the initial radius of all floes.
Wave energy attenuation due to scattering is sufficiently strong in this case that some floes located deep enough in the array do not break.

\begin{figure}
	\begin{center}
		\includegraphics[width=1\textwidth]{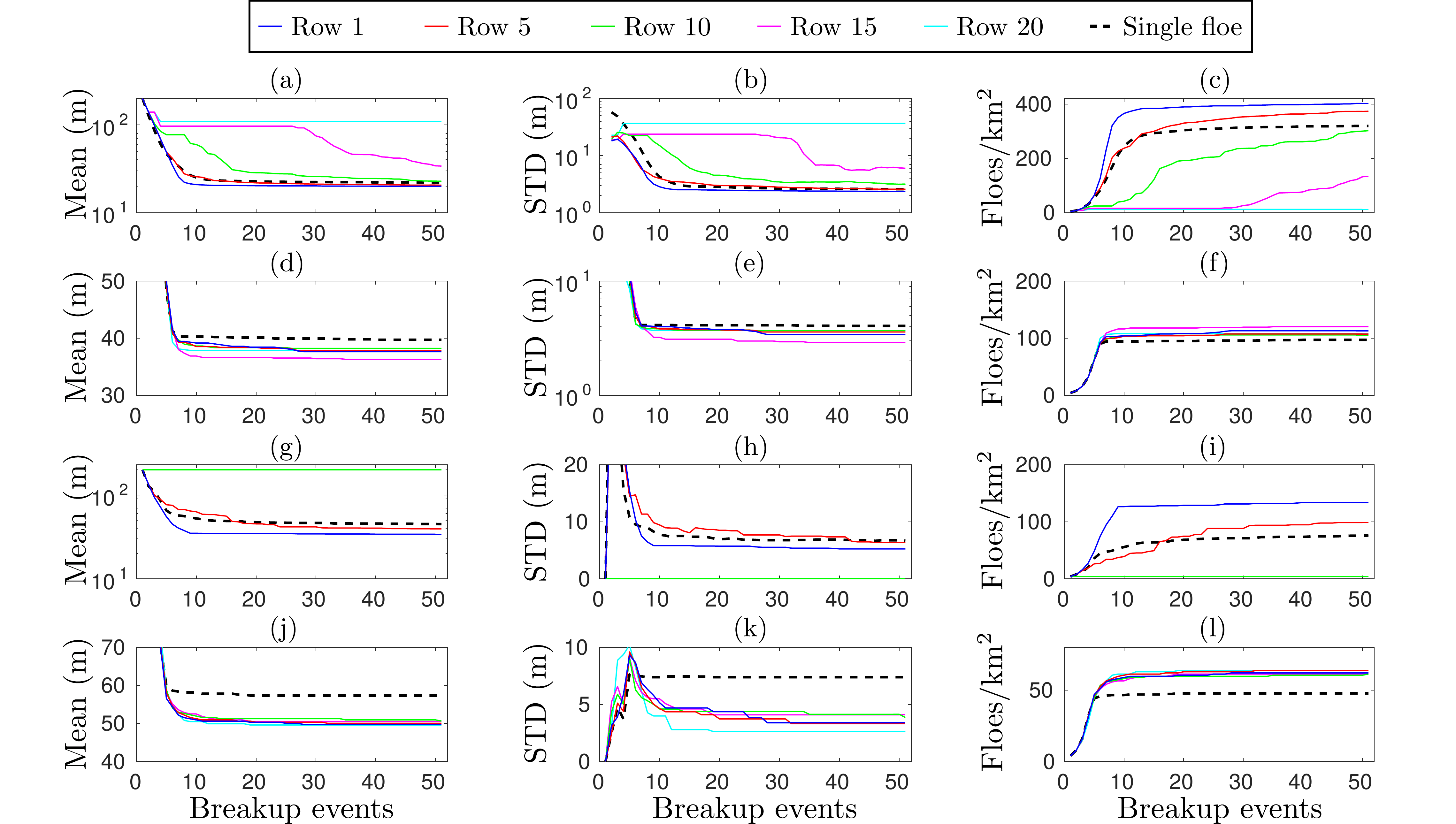}
		\caption{Evolution through 50 breakup events of (a,d,g,j) the mean, (b,e,h,k) standard deviation and (c,f,i,l) number of floes per square kilometre of the FSD in rows 1 (blue), 5 (red), 10 (green), 15 (magenta) and 20 (cyan) of the 20-row array used for simulations with ice thicknesses $D=2$ and 4\,m.
		Results are shown for the four cases (a--c) $D=2\,\text{m}$ and $T=6\,\text{s}$, (d--f) $D=2\,\text{m}$ and $T=14\,\text{s}$, (g--i) $D=4\,\text{m}$ and $T=6\,\text{s}$, and (j--l) $D=4\,\text{m}$ and $T=14\,\text{s}$.
	    The evolution of the FSD statistics obtained for the single floe breakup are shown as black dashed lines for comparison.}
		\label{fig:Mom_multirow_MIZ}
	\end{center}
\end{figure}

To understand the role of scattering on the breakup of the array further, we focus our analysis on the four cases: $(D,T)=(2\,\text{m},6\,\text{s})$, $(2\,\text{m},14\,\text{s})$, $(4\,\text{m},6\,\text{s})$ and $(4\,\text{m},14\,\text{s})$ depicted in figure \ref{fig:StSt_PDFs_MIZ}(d), (f), (g) and (i), respectively.
Figure \ref{fig:Mom_multirow_MIZ} shows the evolution of the mean, standard deviation and number of floes per square kilometre of the FSD in rows 1, 5, 10, 15 and 20 through the 50 breakup events.
For the two long wave cases (i.e.\ $T=14\,\text{s}$) discussed here, we observe that the FSD in all the rows converge to their steady state relatively uniformly. 
Interestingly, enhanced breakup compared to single floe breakup can be seen in all the rows.
The level of breakup differs slightly between the rows, however. 
Inspection of the row-dependence of the FSD steady-state statistics shows an oscillatory behaviour with no clear trend in the case $D=2\,\text{m}$, but with a clear peak in floe number around the 15th row, and a small upward trend in floe number for $D=4\,\text{m}$ suggesting breakup increases with the level of penetration in the array.
The reader is reminded that the results presented in this section are based on a single random realisation of the breakup simulations, so that an ensemble average may be necessary to resolve the small trends associated with the effect of penetration in the array on floe breakup.

For the two short wave cases ($T=6\,\text{s}$) considered in figure \ref{fig:StSt_PDFs_MIZ}, we clearly observe the effect of scattering and associated wave energy attenuation on preventing floe breakup at some distance in the array.
For $D=2\,\text{m}$ (see panels a--c), the FSD in the first row converges quickly to a steady state, while that in row 5 also converges but at a slower rate as the steady state seems to be approximately reached towards the end of the simulation and with a significantly smaller number of floes than in row 1. 
This suggests that wave attenuation has an effect on reducing breakup in the first few rows.
The evolution of the FSD deeper into the array is different, as we observe breakup taking place during the first 4 events but then suddenly stopping.
Breakup then resumes in row 10 and 15 after 8 and 26 events, respectively, while it does not in row 20.
It is hypothesised that large floes in these rows initially fracture, as they do not require much energy to reach the critical breakup stress, as suggested in figure \ref{fig:MCStress}.
After a few breakup events, the floes become too small to break under a wave field strongly attenuated by the front rows.
As breakup in these front rows continues, however, wave attenuation due to scattering by smaller and smaller floes also decreases, so that more wave energy propagates farther into the array with the result that floes are gradually broken up deeper and deeper in the ice field. 
In other words, we have shown in our model that wave-induced breakup has the capacity to reduce the structural integrity of the MIZ, enabling waves to travel farther and cause breakup there, further weakening the ice cover.
This positive feedback process is often used as a motivational concept for observational studies on waves/sea ice interactions.
Note that a steady state is not reached after 50 breakup events in this simulation, so it is possible that breakup starts to resume in row 20 after the rows in front have experienced sufficient breakup.

In the case $D=4\,\text{m}$, breakup in rows 1 and 5 behaves similarly to that for $D=2\,\text{m}$, but rows 10 onward do not experience breakup during the 50 events simulated here.
Because thicker ice tends to increase the degree of wave attenuation, this is a plausible outcome of the simulation.
Inspection of the evolution of the FSD in rows 6, 7, 8 and 9 (not shown here) indicates that breakup in ongoing after 50 events, so that rows 10 onward may gradually start breaking after a larger number of breakup events.

\subsection{Wave spectrum}\label{sec:ResMultiFreqMIZ}

We now consider the breakup of the arrays used in the previous section under a unidirectional wave spectrum, in part acknowledging that the FSD observed in ice-covered oceans is the result of breakup by a sea state composed of a range of frequency and directional components.
Although we cannot reproduce a random sea state in the present model, a numerical experiment is conducted to approximate the breakup induced by a two-parameter Bretschneider spectrum defined by the spectral density function
\begin{equation}
S(T) = \frac{5}{32\pi} T_p H_s^2 \left(\frac{T}{T_p}\right)^5 \expo^{-\frac{5}{4}\left(\frac{T}{T_p}\right)^4},
\label{eq:SpecBret}
\end{equation}
where $T_p$ is the peak wave period and $H_s$ the significant wave height \cite{ochi98}.
Sampling the period spectrum at integer wave periods between 5 and 15\,s, the amplitude of the ambient field at each period is given by
\begin{equation}
A^{\mathrm{am}}(\tau) = \sqrt{2 S(T)}\delta(\tau).
\label{eq:AmpBret}
\end{equation}

For our breakup simulations, we select one wave period $T\,\in\,[5,15]$ randomly at each event and compute the breakup induced by a unidirectional plane wave of period $T$ with amplitude given by \eqref{eq:AmpBret}.
An ensemble of 10 random realisations of the breakup simulation is computed for each ice thickness.
Although this approach is potentially different from breakup by a wave field composed of multiple frequencies, it is conjectured that the randomisation of both wave period and array in conjunction with ensemble averaging provides a legitimate approximation.
We set the parameters of the spectrum to $T_p=10\,\text{s}$ and $H_s=2\,\text{m}$, which corresponds to a typical swell observed in the Southern Ocean \cite{kohout_etal14}.

\begin{figure}
	\begin{center}
		\includegraphics[width=1\textwidth]{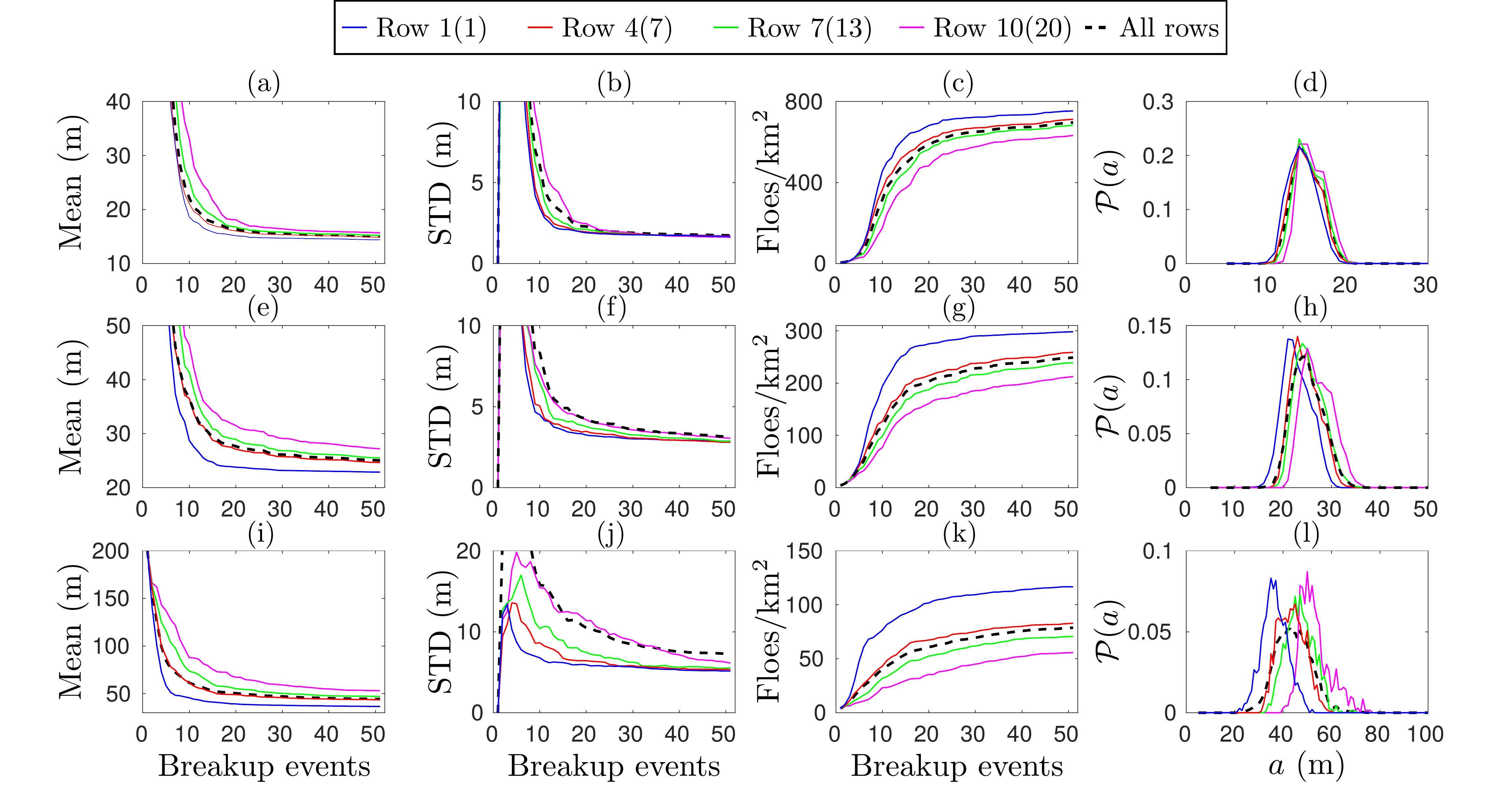}
		\caption{Evolution of (a,e,i) the mean, (b,f,j) the standard deviation and (c,g,k) the number of floes per square kilometre of the FSD, and (d,h,l) probability density functions of the FSD after 50 breakup events.
		Results are shown for the floe thicknesses (a--d) $D=1\,\text{m}$, (e--h) $D=2\,\text{m}$ and (i--l) $D=4\,\text{m}$, and rows 1(1), 4(7), 7(13) and 10(20) in the arrays of floes with $D=1\,\text{m}$ (2 and 4\,m) as solid blue, red, green and magenta lines, respectively.
		We also include results for the entire array (black dashed line) for comparison.
		}
		\label{fig:Stats_multirow_spec_MIZ}
	\end{center}
\end{figure}

Figure \ref{fig:Stats_multirow_spec_MIZ} shows the evolution of the mean, standard deviation and number of floes per square kilometre of the FSD for different rows of the arrays and for the entire array (three leftmost columns of panels).
We observe a clear convergence trend towards a steady state for all rows of the array and for each thickness.
The front rows consistently converge faster than the back ones and to a lower mean and a higher number of floes, so that the degree of breakup decreases with the level of penetration in the array.
Interestingly, the evolution of the mean and number of floes of the FSD for the entire array almost coincides with those associated with the middle row of the array.
This suggests a simple linear dependence of the FSD statistics with respect to the row number, which will be demonstrated later.
The standard deviation of the FSD for the entire array is consitently at least as large as that of the last row, which in turn is the largest of all the rows.
This reflects the larger spread of floe sizes on large scale (entire array) compared to the local scale (each row).

We further display the PDF of the FSD for all cases discussed above in figure \ref{fig:Stats_multirow_spec_MIZ} (rightmost column of panels).
Observe first that the distributions look nearly normal. 
This is likely the result of effectively averaging over the wave periods.
The bi-modality seen in previous distributions does not completely disappear, however, as a shoulder-type feature can be seen on the right of the peak (i.e.\ for large radii), particularly for $D=1$ and 2\,m.
This suggests the existence of a second mode for large radii but with a small peak.

\begin{figure}
	\begin{center}
		\includegraphics[width=1\textwidth]{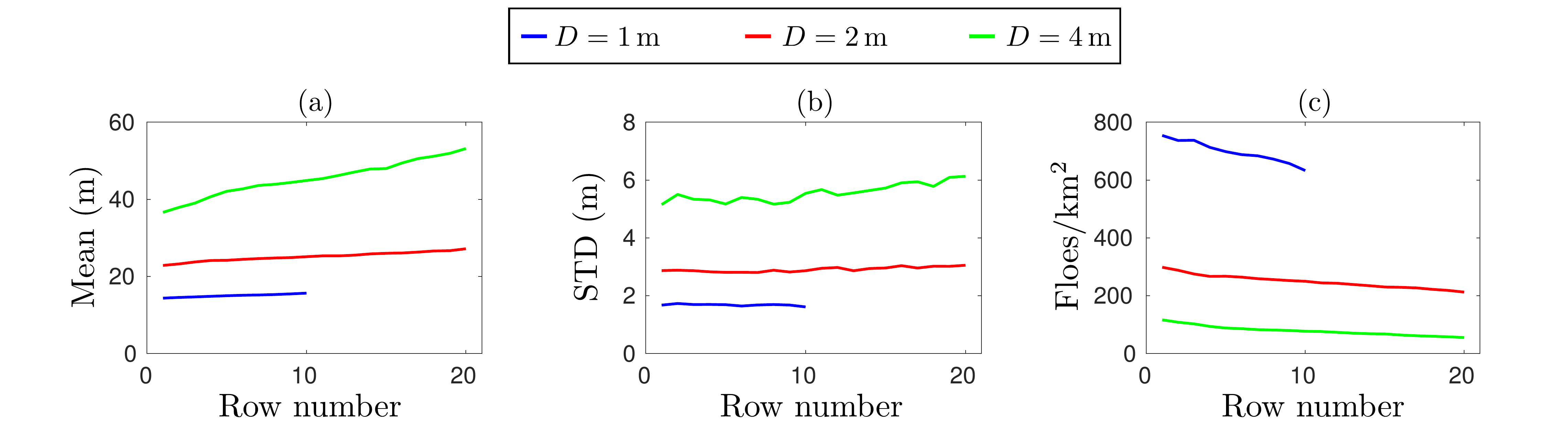}
		\caption{(a) Mean, (b) standard deviation and (c) number of floes per square kilometre after 50 breakup events as a function of row number.}
		\label{fig:Evolrow_multirow_spec_MIZ}
	\end{center}
\end{figure}

The row-dependence of the mean, standard deviation and number of floes after 50 breakup events is analysed further in figure
\ref{fig:Evolrow_multirow_spec_MIZ}.
As indicated above, we observe a linear increase in the mean of the FSD and a linear decrease in the number of floes with respect to row number.
This allows us to estimate the rate of change of mean floe size with respect to penetration in the array, which is important for large scale modelling studies of the MIZ.
Fitting a straight line through the curves generated and extracting the slope, we find the mean floe radius increases at a rate of 0.27, 0.39 and 1.5\,m per kilometre of penetration for $D=1$, 2 and 4\,m, respectively.
This notwithstanding, it is unclear how the limited size of the array affects these estimates.

\section{Conclusion}\label{sec:conclusion}


A new model of ice floe breakup under ocean wave forcing in the MIZ has been developed. 
It combines the time-harmonic multiple scattering theory for a finite array of floating elastic disks proposed by \cite{montiel_etal16} with a parametrisation of flexural failure causing an ice floe to fracture into two floes, provided that the stress field satisfies a particular breakup criterion; the so-called Mohr-Coulomb criterion.
We derived a quantity, referred to as the Mohr-Coulomb stress, that uniquely defines the level of stress at each point of the surface of an ice floe, allowing us to test simply if breakup is expected to take place at any point. 
A numerical experiment was then conducted to analyse the Mohr-Coulomb stress experienced by a single ice floe under a unit amplitude unidirectional wave forcing and determine the regime in which we expect floe breakup.
It was found that 
\begin{enumerate}
	\item a minimum floe diameter exists for each thickness below which breakup cannot occur, and 
	\item this critical diameter depends on wave period in a way that it reaches a minimum at a resonant wave period, for which the floe diameter is approximately equal to the open water wavelength and half the ice-covered wavelength.
\end{enumerate}

A closed-loop feedback algorithm has been proposed to model the evolution of the floe size distribution (FSD) in the MIZ under a sustained wave event.
Each loop consists of (i)~computing the Mohr-Coulomb stress in all the floes, (ii)~breaking up each floe satisfying the Mohr-Coulomb criterion and (iii)~generating a new array of floes from the updated FSD, which is then used as the geometry of the wave interaction problem in the next loop.
We conducted a number of numerical experiments to determine the evolution of the FSD towards a steady state through 50 breakup events (i.e.\ loops), for different wave and ice configurations.
We simulated the breakup of (i)~a single large floe for a monochromatic unit-amplitude plane wave forcing, (ii)~an array of large floes for the same monochromatic forcing, and (iii) an array of large floes for a Bretschneider spectrum forcing.
Key findings are summarised below.
\begin{enumerate}
	\item Breakup of a single large floe causes the emergence of a bi-modal FSD for most wave and ice parameters considered.  
	Larger values of wave period and ice thickness correspond to FSDs with larger floe sizes and more separated peaks of the associated bi-modal distribution. 
	The convergence of the FSD towards its steady state under repeated breakup events is very quick for long waves and slower for short waves.
	Increasing values of ice thickness also tends to decrease the rate of convergence of the FSD, suggesting multiple wave scattering within the array of broken floes enhances breakup and therefore influences the steady state FSD.
	
	\item Breakup of an array of large floes under monochromatic forcing provides additional insight into the effect of multiple scattering on the FSD.
	First, the bi-modality of the FSD observed for the single floe breakup simulations is either damped or removed, as the population of large floes associated with the second mode is consistently redistributed to smaller floes.
	Second, results of our simulations indicate two scattering regimes exist, i.e.\ long waves/thin ice and short waves/thick ice. 
	In the former regime, wave-induced ice breakup is enhanced compared to the single floe breakup, which is likely to be a consequence of constructive interference of wave fields radiated by the individual floes.
	In the second regime, multiple scattering causes sufficient wave energy attenuation through the array to prevent some ice floes from fracturing, resulting in broader FSDs compared to those obtained from the repeated breakup of a single floe.
	
	\item Investigation of the evolution of the FSD at different levels of penetration in the array indicates that enhanced floe breakup in the long wave regime occurs throughout the array with a small upward trend with distance from the ice edge.
	In the short wave regime, we observe the positive feedback between wave-induced ice breakup and ice-induced wave attenuation.
	Breakup originally only takes place in the front rows as waves are attenuated by scattering deeper in the array.
	After sufficient breakup, waves are less attenuated, carry energy at a higher level of penetration and cause breakup there. 
	The overall effect is the observation of a breakup front marching forward in the MIZ as the structural integrity of the ice cover reduces under wave-induced breakup.
	
	\item Breakup of an array under a Bretschneider spectrum forcing generates near-normal FSDs for all ice thicknesses, which is likely to be a consequence of averaging with respect to wave period.
	As opposed to the simulations with monochromatic forcing, breakup decreases with the level of penetration in the array, such that the mean floe size increases linearly with distance from the ice edge.
	The rate of floe size increase is larger for thicker ice.	
\end{enumerate}

An important outcome of our investigation is that no power-law FSD was generated from the simulations.
The number of floes consistently decreases to zero for smaller floe sizes.
This results from the fact that small floes are less prone to elastic deformations than large floes, i.e.\ they behave similarly to a rigid body. 
Analysis of the data generated in \textsection\,\ref{sec:BrkpCrit}\ref{sec:BrkpStress_1floe} shows that the wave energy required to cause breakup in small floes increases exponentially fast as floe size decreases below a critical size, so that flexural failure by ocean waves cannot be responsible for the observed increasing number of small floes in the MIZ \cite{toyota_etal11}.
Our analysis suggests that wave-induced floe breakup does not create or preserve the observed power-law FSD.
Other processes acting on longer time scales (e.g.\ thermodynamics or collisions) must be considered to explain this feature of the FSD.
Although recent modelling work has shown the emergence of a power-law FSD \cite{horvat_tziperman15,zhang_etal15,zhang_etal16}, it is still unclear how each process contributes to the observed result.
It should also be noted that we attempted to fit a power-law curve in the large floe regime of the FSDs obtained from our simulations, i.e.\ for radii larger than the peak radius, but the limited extent of this regime (i.e.\ spanning less than one order of magnitude in floe radius) did not allow us to obtain statistically significant results.

Although our findings provide much theoretical understanding of the wave-induced ice breakup process in the MIZ, the underlying model was constructed based on a number simplifying assumptions, which may influence certain results.
Specifically, the validity of our breakup model, which involves breaking a circular floe into two circular floes and time-harmonic wave forcing, is unclear and requires further investigation. 
It would be difficult to relax these assumptions in the context of the three-dimensional wave scattering model considered here.
We may envisage a simpler two-dimensional model, however, in which the one-dimensional ice cover is initialised as a semi-infinite beam and forced by a transient incident wave generated from a frequency spectrum.
Monitoring the evolution of the FSD in an area of ice-covered ocean during a wave breakup event either in the field or in a controlled laboratory setting, would be needed to provide a clearer picture of the complicated processes investigated here.

\vskip6pt

\enlargethispage{20pt}

\noindent \textbf{Data Accessibility}. The Matlab codes and data files created for the numerical investigation reported here can be accessed through:\\ \textbf{http://www.maths.otago.ac.nz/files/icebreakup/Data\_breakup.zip}. \\
\noindent \textbf{Authors' Contributions}. FM and VAS both developed the model conceptually. VAS proposed using the Mohr-Coulomb failure criterion for flexural breakup of sea ice. FM conducted the investigation and drafted the manuscript. VAS helped editing the manuscript. Both authors gave final approval for submission. \\
\noindent \textbf{Competing Interests}. We declare we have no competing interests. \\
\noindent \textbf{Funding}. The authors are indebted to ONR Departmental Research Initiative `Sea State and Boundary Layer Physics of the Emerging Arctic Ocean' (award number N00014-131-0279), EU FP7 Grant (SPA-2013.1.1-06) and the University of Otago, for their support.


\bibliographystyle{vancouver}

\end{document}